# Computation of Optimal Control Problems with Terminal Constraint via Modified Evolution Partial Differential Equation

Sheng ZHANG, Kai-Feng HE, and Fei LIAO

*Abstract:* **The Variation Evolving Method (VEM), which seeks the optimal solutions with the variation evolution principle, is further developed to be more flexible in solving the Optimal Control Problems (OCPs) with terminal constraint. With the first-order stable dynamics to eliminate the infeasibilities, the Modified Evolution Partial Differential Equation (MEPDE) that is valid in the infeasible solution domain is proposed, and a Lyapunov functional is constructed to theoretically ensure its validity. In particular, it is proved that even with the infinite-time convergence dynamics, the violated terminal inequality constraints, which are inactive for the optimal solution, will enter the feasible domain in finite time. Through transforming the MEPDE to the finite-dimensional Initial-value Problem (IVP) with the semi-discrete method, the OCPs may be solved with common Ordinary Differential Equation (ODE) numerical integration methods. Illustrative examples are presented to show the effectiveness of the proposed method.**

## I. Introduction

Optimal control theory aims to determine the inputs to a dynamic system that optimize a specified performance index while satisfying constraints on the motion of the system. It is closely related to engineering and has been widely studied [1]. Because of the complexity, Optimal Control Problems (OCPs) are usually solved with numerical methods. Various numerical methods are developed and generally they are divided into two classes, namely, the direct methods and the indirect methods [2]. The direct methods discretize the control or/and state variables to obtain the Nonlinear Programming (NLP) problem, for example, the widely-used direct shooting method [3] and the classic collocation method [4]. These methods are easy to apply, whereas the results obtained are usually suboptimal [5], and the optimal may be infinitely approached. The indirect methods transform the OCP to a Boundary-value Problem (BVP) through the optimality conditions. Typical methods of this type include the well-known indirect shooting method [2] and the novel symplectic method [6]. Although be more precise, the indirect methods often suffer from the significant numerical difficulty due to the ill-conditioning of the Hamiltonian dynamics, that is, the stability of costates dynamics is adverse to that of the states dynamics [7]. The recent development, representatively the Pseudo-spectral (PS) method [8], blends the two types of methods, as it unifies the NLP and the BVP in a dualization view [9]. Such methods inherit the advantages of both types and blur their difference.

The authors are with the Computational Aerodynamics Institution, China Aerodynamics Research and Development Center, Mianyang, 621000, China. (e-mail: zszhangshengzs@hotmail.com).

Theories in the control field often enlighten strategies for the optimal control computation, for example, the non-linear variable transformation to reduce the variables [10]. Recently, a Variation Evolving Method (VEM), which is enlightened by the states evolution within the stable continuous-time dynamic system, is proposed for the optimal control computation [11]-[16]. The VEM also synthesizes the direct and indirect methods, but from a new standpoint. The Evolution Partial Differential Equation (EPDE), which describes the evolution of variables towards the optimal solution, is derived from the viewpoint of variation motion, and the optimality conditions will be gradually met under this frame. In Refs. [11] and [12], besides the states and the controls, the costates are also employed in developing the EPDE, and this increases the complexity of the computation. In Ref. [13], a compact version of the VEM that uses only the original variables is proposed. The costate-free optimality conditions are established and the corresponding EPDE is derived. However, in that work, only a class of OCPs with free terminal states is handled. In Refs. [14] and [15], the compact VEM is furthered developed to address the OCPs with terminal Equality Constraints (ECs) and Inequality Constraints (IECs). Normally, under the frame of the compact VEM, the definite conditions for the EPDE are required to be feasible solutions, and this is inflexible for application. In Ref. [16], the Modified EPDE (MEPDE) that uses arbitrary definite conditions but still seeks the optimal solution is proposed to facilitate the computation of the OCPs without terminal constraint. In this paper, it is further developed to accommodate the OCPs with terminal constraint.

Throughout the paper, our work is built upon the assumption that the solution for the optimization problem exists. We do not describe the existing conditions for the purpose of brevity. Relevant researches such as the Filippov-Cesari theorem are documented in [17]. In the following, first the principles of VEM and the MEPDE for OCPs with free terminal states are briefly reviewed. Then the MEPDE that accommodates arbitrary definite conditions for the OCPs with terminal constraints are developed. Later illustrative examples are solved to verify the effectiveness of the equations.

## II. Preliminaries

### A. Principle of VEM

The VEM is a newly developed method for the optimal solutions. It is enlightened from the inverse consideration of the Lyapunov dynamics stability theory in the control field [18]. As the start point of this method, the generalized Lyapunov principle for the infinite-dimensional stable continuous-time dynamics may be stated as

**Lemma 1**: For an infinite-dimensional dynamic system described by

$$\frac{\delta \boldsymbol{y}(x)}{\delta t} = \boldsymbol{f}(\boldsymbol{y}, x) \tag{1}$$

or presented equivalently in the Partial Differential Equation (PDE) form as

$$\frac{\partial \boldsymbol{y}(x,t)}{\partial t} = \boldsymbol{f}(\boldsymbol{y}, x) \tag{2}$$

where “$\delta$” denotes the variation operator and “$\partial$” denotes the partial differential operator. $t \in \mathbb{R}$ is the time. $x \in \mathbb{R}$ is the independent variable, $\boldsymbol{y}(x) \in \mathbb{R}^n(x)$ is the function vector of $x$, and $\boldsymbol{f} : \mathbb{R}^n(x) \times \mathbb{R} \to \mathbb{R}^n(x)$ is a vector function. Let $\hat{\boldsymbol{y}}(x)$, contained within a certain function set $\mathbb{D}(x)$, is an equilibrium function that satisfies $\boldsymbol{f}(\hat{\boldsymbol{y}}(x), x) = \boldsymbol{0}$. If there exists a continuously differentiable (If not, only except at $\hat{\boldsymbol{y}}(x)$) functional $V : \mathbb{D}(x) \to \mathbb{R}$ such that

i) $V\left(\hat{\boldsymbol{y}}(x)\right) = c$ and $V\left(\boldsymbol{y}(x)\right) > c$ in $\mathbb{D}(x) / \{\hat{\boldsymbol{y}}(x)\}$.

ii) $\frac{\delta}{\delta t} V\left(\boldsymbol{y}(x)\right) \le 0$ in $\mathbb{D}(x)$ and $\frac{\delta}{\delta t} V\left(\boldsymbol{y}(x)\right) < 0$ in $\mathbb{D}(x) / \{\hat{\boldsymbol{y}}(x)\}$.

where $c$ is a constant. Then $\boldsymbol{y}(x)=\hat{\boldsymbol{y}}(x)$ is an asymptotically stable solution in $\mathbb{D}(x)$.

The VEM analogizes the optimal solution to the asymptotically stable equilibrium point of an infinite-dimensional dynamic system, and derives such dynamics that minimize the specific performance index (act as the Lyapunov functional) within an optimization problem. To implement the idea, a virtual dimension, the variation time $\tau$, is introduced to describe the process that a variable $\boldsymbol{x}(t)$ evolves to the optimal solution to minimize the performance index under the dynamics governed by the variation dynamic evolution equations (in the form of Eq. (1)). Fig. 1 illustrates the variation evolution of variables in the VEM to solve the OCP. Through the variation motion, the initial guess of variables will evolve to the optimal solution.

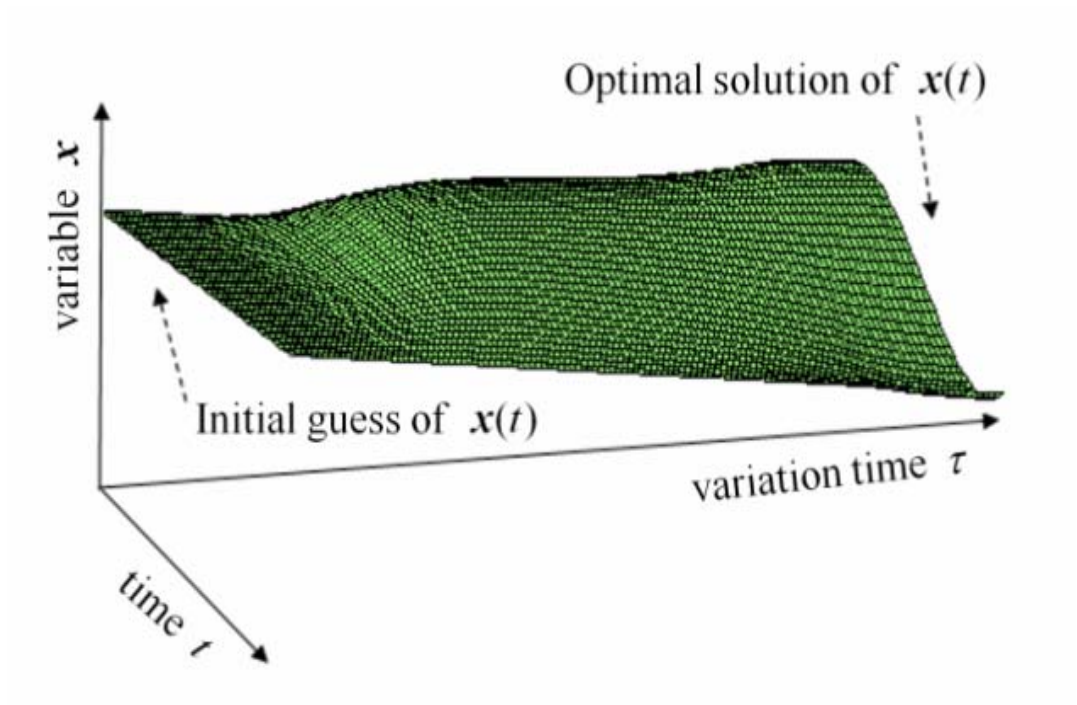

Fig. 1. The illustration of the variable evolving along the variation time $\tau$ in the VEM.

The VEM is first demonstrated for the unconstrained calculus-of-variations problems, and then applied to typical OCPs [11][13]. The variation dynamic evolution equations, derived under the frame of the VEM, may be reformulated as the EPDE and the Evolution Differential Equation (EDE), by replacing the variation operator "$\delta$" with the partial differential operator "$\partial$" and the differential operator "$\mathrm{d}$". Under the dynamics governed by the EPDE, the variables will achieve the optimality conditions gradually. For example, consider the calculus-of-variations problems defined as

$$J=\int_{t_0}^{t_f} F\left(\boldsymbol{y}(t),\dot{\boldsymbol{y}}(t),t\right)\mathrm{d}t \tag{3}$$

where the elements of the variable vector $\boldsymbol{y}(t)\in\mathbb{R}^n(t)$ belong to $C^2[t_0,t_f]$. $t_0$ and $t_f$ are the fixed initial and terminal time, and the boundary conditions are prescribed as $\boldsymbol{y}(t_0)=\boldsymbol{y}_0$ and $\boldsymbol{y}(t_f)=\boldsymbol{y}_f$. The variation dynamic evolution equation obtained with the VEM is

$$\frac{\delta \boldsymbol{y}}{\delta \tau}=-\boldsymbol{K}\left(F_{\boldsymbol{y}}-\frac{\mathrm{d}}{\mathrm{d}t}(F_{\dot{\boldsymbol{y}}})\right) \tag{4}$$

where the column vectors $F_{\boldsymbol{y}}=\frac{\partial F}{\partial \boldsymbol{y}}$ and $F_{\dot{\boldsymbol{y}}}=\frac{\partial F}{\partial \dot{\boldsymbol{y}}}$ are the shorthand notations of partial derivatives, and $\boldsymbol{K}$ is a $n\times n$ dimensional positive-definite matrix. Correspondingly, the reformulated EPDE is

$$\frac{\partial \boldsymbol{y}}{\partial \tau}=-\boldsymbol{K}\left(F_{\boldsymbol{y}}-\frac{\partial}{\partial t}(F_{\dot{\boldsymbol{y}}})\right) \tag{5}$$

The equilibrium solution of EPDE (5) will satisfy the optimality condition, i.e., the Euler-Lagrange equation [19][20]

$$F_{\boldsymbol{y}}-\frac{\mathrm{d}}{\mathrm{d}t}(F_{\dot{\boldsymbol{y}}})=\boldsymbol{0} \tag{6}$$

Since the right function of the EPDE only depends on the time $t$, it is suitable to be solved with the well-known semi-discrete method in the field of PDE numerical calculation [21]. Through the discretization along the normal time dimension, the EPDE is transformed to the finite-dimensional Initial-value Problem (IVP) to be solved, with the common Ordinary Differential Equation (ODE) integration methods. Note that the resulting IVP is defined with respect to the variation time $\tau$ , not the normal time $t$ .

*B. The MEPDE*

Normally, the EPDE derived under the frame of the compact VEM is only valid within the feasible solution domain, in which the solutions satisfy various constraints including the dynamics constraint and the boundary conditions. The definite conditions of the EPDE for an OCP, i.e., $\left.\begin{bmatrix}\boldsymbol{x}(t,\tau)\\ \boldsymbol{u}(t,\tau)\end{bmatrix}\right|_{\tau=0}$ , are required to be feasible. Using the EPDE formulation to achieve the evolution process as depicted in Fig. 1, the feasible initial conditions of $\boldsymbol{x}(t,\tau)$ and $\boldsymbol{u}(t,\tau)$ at $\tau=0$ needs to be determined first and they will reach the optimal solution of the OCP at $\tau=+\infty$ .

However, for the OCPs with terminal constraints, finding a feasible solution is not an easy task. In Ref. [16], the EPDE is modified to accommodate definite conditions that are infeasible in the dynamics and initial boundary conditions. The basic principle that turns an infeasible solution to be feasible is the asymptotically stability of first-order dynamic systems, that is, an error variable $e(t)$ (and a parameter trivially) will be driven to be zero in terms of the following equation

$$\frac{\delta e(t)}{\delta\tau}=-ke(t) \tag{7}$$

or equivalently written as

$$\frac{\partial e(t,\tau)}{\partial\tau}=-ke(t) \tag{8}$$

where $k$ is a positive constant.

For the OCPs with free terminal states, Eq. (8) is used to develop the following MEPDE as

$$\frac{\partial\boldsymbol{x}}{\partial\tau}=-\boldsymbol{\Phi}_o(t,t_0)\boldsymbol{K}_{x_0}\left(\boldsymbol{x}(t_0)-\boldsymbol{x}_0\right)+\int_{t_0}^{t}\boldsymbol{\Phi}_o(t,s)\boldsymbol{f}_{\boldsymbol{u}}(s)\frac{\partial\boldsymbol{u}}{\partial\tau}(s)\,\mathrm{d}s-\int_{t_0}^{t}\boldsymbol{\Phi}_o(t,s)\boldsymbol{K}_f\left(\frac{\partial\boldsymbol{x}}{\partial t}(s)-\boldsymbol{f}(s)\right)\mathrm{d}s \tag{9}$$

where $\boldsymbol{K}_{x_0}$ and $\boldsymbol{K}_f$ are positive-definite matrixes. $\boldsymbol{\Phi}_o(t,s)$ is the state transition matrix from time point $s$ to time point $t$ , which satisfies

$$\frac{\partial}{\partial t}\boldsymbol{\Phi}_o(t,s)=\boldsymbol{f}_{\boldsymbol{x}}(t)\boldsymbol{\Phi}_o(t,s) \tag{10}$$

$\boldsymbol{f}_{\boldsymbol{x}}$ and $\boldsymbol{f}_{\boldsymbol{u}}$ are the Jacobi matrixes of the state dynamics right function $\boldsymbol{f}(\boldsymbol{x},\boldsymbol{u},t)$ . $\boldsymbol{x}_0$ is the prescribed initial boundary condition. Eq. (9) implies that (See Ref. [16])

$$\frac{\mathrm{d}\boldsymbol{x}(t_0)}{\mathrm{d}\tau}=-\boldsymbol{K}_{x_0}\left(\boldsymbol{x}(t_0)-\boldsymbol{x}_0\right) \tag{11}$$

and

$$\frac{\partial}{\partial\tau}\left(\frac{\partial\boldsymbol{x}}{\partial t}-\boldsymbol{f}\right)=-\boldsymbol{K}_f\left(\frac{\partial\boldsymbol{x}}{\partial t}-\boldsymbol{f}\right) \tag{12}$$

This means during the evolution process, the violations on the initial boundary conditions and the dynamics constraints will be gradually eliminated. More than being intuitive, with the MEPDE (9), the evolution towards to the optimal solution is theoretically guaranteed by an unconstrained Lyapunov functional.

## III. MEPDE FOR OCPS WITH TERMINAL CONSTRAINTS

### A. Problem definition

In this paper, we consider the OCPs with terminal constraints that are defined as

**Problem 1**: Consider performance index of Bolza form

$$J = \varphi(\boldsymbol{x}(t_f), t_f) + \int_{t_0}^{t_f} L\left(\boldsymbol{x}(t), \boldsymbol{u}(t), t\right) \mathrm{d}t \tag{13}$$

subject to the dynamic equation

$$\dot{\boldsymbol{x}} = \boldsymbol{f}(\boldsymbol{x}, \boldsymbol{u}, t) \tag{14}$$

where $t \in \mathbb{R}$ is the time. $\boldsymbol{x} \in \mathbb{R}^n$ is the state vector and its elements belong to $C^2[t_0, t_f]$. $\boldsymbol{u} \in \mathbb{R}^m$ is the control vector and its elements belong to $C^1[t_0, t_f]$. The function $L: \mathbb{R}^n \times \mathbb{R}^m \times \mathbb{R} \to \mathbb{R}$ and its first-order partial derivatives are continuous with respect to $\boldsymbol{x}$, $\boldsymbol{u}$ and $t$. The function $\varphi: \mathbb{R}^m \times \mathbb{R} \to \mathbb{R}$ and its first-order and second-order partial derivatives are continuous with respect to $\boldsymbol{x}$ and $t$. The vector function $\boldsymbol{f}: \mathbb{R}^n \times \mathbb{R}^m \times \mathbb{R} \to \mathbb{R}^n$ and its first-order partial derivatives are continuous and Lipschitz in $\boldsymbol{x}$, $\boldsymbol{u}$ and $t$. The initial time $t_0$ is fixed and the terminal time $t_f$ is free. The initial and terminal boundary conditions are respectively prescribed as

$$\boldsymbol{x}(t_0) = \boldsymbol{x}_0 \tag{15}$$

$$\boldsymbol{g}_E\left(\boldsymbol{x}(t_f), t_f\right) = \boldsymbol{0} \tag{16}$$

$$\boldsymbol{g}_I\left(\boldsymbol{x}(t_f), t_f\right) \leq \boldsymbol{0} \tag{17}$$

where $\boldsymbol{g}_E: \mathbb{R}^n \times \mathbb{R} \to \mathbb{R}^{q_E}$ is a $q_E$ dimensional vector function with continuous first-order partial derivatives, and $\boldsymbol{g}_I: \mathbb{R}^n \times \mathbb{R} \to \mathbb{R}^{q_I}$ is a $q_I$ dimensional vector function with continuous first-order partial derivatives. Find the optimal solution $(\hat{\boldsymbol{x}}, \hat{\boldsymbol{u}})$ that minimizes $J$, i.e.

$$(\hat{\boldsymbol{x}}, \hat{\boldsymbol{u}}) = \arg\min(J) \tag{18}$$

### B. Evolution equations in feasible solution domain

In Ref. [15], we have derived the variation dynamic evolution equations for Problem 1 within the feasible solution domain $\mathbb{D}_o$, namely

$$\frac{\delta \boldsymbol{x}}{\delta \tau} = \int_{t_0}^{t} \boldsymbol{\Phi}_o(t, s) \boldsymbol{f}_{\boldsymbol{u}}(s) \frac{\delta \boldsymbol{u}}{\delta \tau}(s) \mathrm{d}s \tag{19}$$

$$\frac{\delta \boldsymbol{u}}{\delta \tau} = -\boldsymbol{K} \boldsymbol{p}_{\boldsymbol{u}}^{tc} \tag{20}$$

$$\frac{\delta t_f}{\delta \tau} = -k_{t_f} p_{t_f}^{tc} \tag{21}$$

where $\boldsymbol{K}$ is the $m \times m$ dimensional positive-definite matrix and $k_{t_f}$ is a positive constant.

$$\boldsymbol{p}_{\boldsymbol{u}}^{tc} = \boldsymbol{p}_{\boldsymbol{u}} + \boldsymbol{f}_{\boldsymbol{u}}^{\mathrm{T}} \boldsymbol{\Phi}_o^{\mathrm{T}}(t_f, t)\left((\boldsymbol{g}_E)_{\boldsymbol{x}_f}{}^{\mathrm{T}} \boldsymbol{\pi}_E + (\boldsymbol{g}_I)_{\boldsymbol{x}_f}{}^{\mathrm{T}} \boldsymbol{\pi}_I\right) \tag{22}$$

$$\boldsymbol{p}_{\boldsymbol{u}}(t) = L_{\boldsymbol{u}} + \boldsymbol{f}_u{}^{\mathrm{T}} \varphi_{\boldsymbol{x}} + \boldsymbol{f}_u{}^{\mathrm{T}}\left(\int_t^{t_f} \boldsymbol{\Phi}_o{}^{\mathrm{T}}(\sigma, t)\left(L_{\boldsymbol{x}}(\sigma) + \varphi_{t\boldsymbol{x}}(\sigma) + \varphi_{\boldsymbol{xx}}{}^{\mathrm{T}}(\sigma) \boldsymbol{f}(\sigma) + \boldsymbol{f}_{\boldsymbol{x}}(\sigma)^{\mathrm{T}} \varphi_{\boldsymbol{x}}(\sigma)\right) \mathrm{d}\sigma\right) \tag{23}$$

$$p_{t_f}^{tc} = \left(\varphi_t + {\varphi_{\boldsymbol{x}}}^{\mathrm{T}} \boldsymbol{f} + L + {\boldsymbol{\pi}_E}^{\mathrm{T}} \left((\boldsymbol{g}_E)_{\boldsymbol{x}_f} \boldsymbol{f} + (\boldsymbol{g}_E)_{t_f}\right) + {\boldsymbol{\pi}_I}^{\mathrm{T}} \left((\boldsymbol{g}_I)_{\boldsymbol{x}_f} \boldsymbol{f} + (\boldsymbol{g}_I)_{t_f}\right)\right)\Big|_{t_f} \tag{24}$$

The parameter vectors $\boldsymbol{\pi}_E \in \mathbb{R}^{q_E}$ and $\boldsymbol{\pi}_I \in \mathbb{R}^{q_I}$ are determined by

$$\boldsymbol{\pi}_E = \begin{bmatrix} \pi_1 \\ \pi_2 \\ \cdots \\ \pi_{q_E} \end{bmatrix}$$
$$(\pi_I)_i = 0 \qquad i \notin \mathbb{I}_p, \ \boldsymbol{\pi}_I(\mathbb{I}_p) = \begin{bmatrix} \pi_{q_E+1} \\ \pi_{q_E+2} \\ \cdots \\ \pi_{q_E+n_{\mathbb{I}_p}} \end{bmatrix} \geq \boldsymbol{0} \tag{25}$$

where the index set $\mathbb{I}_p$ is defined as

$$\mathbb{I}_p = \{ i \,|\, (g_I)_i = 0, \ \frac{\delta (g_I)_i}{\delta \tau} \leq 0 \text{ is an active IEC}, \ i = 1, 2, ..., q_I \} \tag{26}$$

and its number is denoted by $n_{\mathbb{I}_p}$. Note that the active IEC is defined for the Feasibility-preserving Evolution Optimization Problem (FPEOP) [15]. The parameter $\boldsymbol{\pi} \in \mathbb{R}^{q_E + n_{\mathbb{I}_p}}$ is the solution of the linear matrix equation

$$\boldsymbol{M}\boldsymbol{\pi} = -\boldsymbol{r} \tag{27}$$

and the $(q_E + n_{\mathbb{I}_p}) \times (q_E + n_{\mathbb{I}_p})$ dimensional matrix $\boldsymbol{M}$ and the $q_E + n_{\mathbb{I}_p}$ dimensional vector $\boldsymbol{r}$ are

$$\boldsymbol{M} = \boldsymbol{g}_{\boldsymbol{x}_f} \left( \int_{t_0}^{t_f} \boldsymbol{\Phi}_o(t_f, t) \boldsymbol{f}_{\boldsymbol{u}} \boldsymbol{K} {\boldsymbol{f}_{\boldsymbol{u}}}^{\mathrm{T}} {\boldsymbol{\Phi}_o}^{\mathrm{T}}(t_f, t) \, \mathrm{d}t \right) {\boldsymbol{g}_{\boldsymbol{x}_f}}^{\mathrm{T}} + k_{t_f} (\boldsymbol{g}_{\boldsymbol{x}_f} \boldsymbol{f} + \boldsymbol{g}_{t_f})(\boldsymbol{g}_{\boldsymbol{x}_f} \boldsymbol{f} + \boldsymbol{g}_{t_f})^{\mathrm{T}} \Big|_{t_f} \tag{28}$$

$$\boldsymbol{r} = \boldsymbol{g}_{\boldsymbol{x}_f} \left( \int_{t_0}^{t_f} \boldsymbol{\Phi}_o(t_f, t) \boldsymbol{f}_{\boldsymbol{u}} \boldsymbol{K} \boldsymbol{p}_{\boldsymbol{u}} \, \mathrm{d}t \right) + k_{t_f} (\boldsymbol{g}_{\boldsymbol{x}_f} \boldsymbol{f} + \boldsymbol{g}_{t_f}) (\varphi_t + {\varphi_{\boldsymbol{x}}}^{\mathrm{T}} \boldsymbol{f} + L) \Big|_{t_f} \tag{29}$$

with $\boldsymbol{g} = \begin{bmatrix} \boldsymbol{g}_E \\ \boldsymbol{g}_I(\mathbb{I}_p) \end{bmatrix}$.

Starting from a feasible solution, the variables under the evolution equations (19), (20), and (21) maintain the feasibility conditions (14)-(17) and will ultimately satisfy the costate-free optimality conditions

$$\boldsymbol{p}_{\boldsymbol{u}}^{tc} = \boldsymbol{0} \tag{30}$$

$$p_{t_f}^{tc} = 0 \tag{31}$$

Correspondingly, the index set $\mathbb{I}_p$ will evolve to $\hat{\mathbb{I}}_p$, which denotes the index set for the optimal solution. In particular, the relation to the costates in the classic optimality conditions is uncovered as

$$\boldsymbol{\lambda}(t) = \varphi_{\boldsymbol{x}} + {\boldsymbol{\Phi}_o}^{\mathrm{T}}(t_f, t) \left( {(\boldsymbol{g}_E)_{\boldsymbol{x}_f}}^{\mathrm{T}} \boldsymbol{\pi}_E + {(\boldsymbol{g}_I)_{\boldsymbol{x}_f}}^{\mathrm{T}} \boldsymbol{\pi}_I \right) + \int_t^{t_f} {\boldsymbol{\Phi}_o}^{\mathrm{T}}(\sigma, t) \left( L_{\boldsymbol{x}}(\sigma) + \varphi_{t\boldsymbol{x}}(\sigma) + {\varphi_{\boldsymbol{x}\boldsymbol{x}}}^{\mathrm{T}}(\sigma) \boldsymbol{f}(\sigma) + \boldsymbol{f}_{\boldsymbol{x}}(\sigma)^{\mathrm{T}} \varphi_{\boldsymbol{x}}(\sigma) \right) \mathrm{d}\sigma \tag{32}$$

where $\boldsymbol{\pi}_E$ and $\boldsymbol{\pi}_I$ correspond to the Lagrange multipliers and Karush-Kuhn-Tucker (KKT) multipliers that adjoin the terminal ECs and IECs respectively.

*C. Modification of variation dynamic evolution equations*

To increase the flexibility of computation, we hope that the evolution equations may be valid in the infeasible solution domain, and thereby a feasible initial solution is not required. In this sub-section, we will address the problem of driving an infeasible solution that violates Eqs. (14)-(17) to be feasible, and Eq. (8) is still the basic principle that we resort to. At first, we again employ Eq. (9) to eliminate the errors in the dynamics constraint (14) and the initial boundary conditions (15), and its variation form is

$$\frac{\delta \boldsymbol{x}(t)}{\delta \tau} = -\boldsymbol{\Phi}_o(t,t_0)\boldsymbol{K}_{\boldsymbol{x}_0}\boldsymbol{e}_{\boldsymbol{x}_0} + \int_{t_0}^{t} \boldsymbol{\Phi}_o(t,s)\boldsymbol{f}_{\boldsymbol{u}}(s)\frac{\delta \boldsymbol{u}}{\delta \tau}(s)\,\mathrm{d}s - \int_{t_0}^{t} \boldsymbol{\Phi}_o(t,s)\boldsymbol{K}_{\boldsymbol{f}}\boldsymbol{e}_{\boldsymbol{f}}\,\mathrm{d}s \tag{33}$$

where $\boldsymbol{e}_{\boldsymbol{f}}(t) = \dot{\boldsymbol{x}} - \boldsymbol{f}(\boldsymbol{x},\boldsymbol{u},t)$ is the dynamics error variable and $\boldsymbol{e}_{\boldsymbol{x}_0} = \boldsymbol{x}(t_0) - \boldsymbol{x}_0$ is the initial state error. Before we focus on treating the terminal constraints, it is worthwhile to investigate how the performance index given by Eq. (13) varies when starting from an infeasible solution. See

$$\begin{aligned}\frac{\delta J}{\delta \tau} &= \frac{\delta}{\delta \tau}\left(\int_{t_0}^{t_f}\left(\varphi_t + \varphi_{\boldsymbol{x}}{}^{\mathrm{T}}\dot{\boldsymbol{x}} + L(\boldsymbol{x},\boldsymbol{u},t)\right)\mathrm{d}t\right) \\ &= (\varphi_t + \varphi_{\boldsymbol{x}}{}^{\mathrm{T}}\dot{\boldsymbol{x}} + L)\Big|_{t_f}\frac{\delta t_f}{\delta \tau} + \int_{t_0}^{t_f}\left((\varphi_{t\boldsymbol{x}}{}^{\mathrm{T}} + \dot{\boldsymbol{x}}^{\mathrm{T}}\varphi_{\boldsymbol{xx}} + \varphi_{\boldsymbol{x}}{}^{\mathrm{T}}\boldsymbol{f}_{\boldsymbol{x}} + L_{\boldsymbol{x}}{}^{\mathrm{T}})\frac{\delta \boldsymbol{x}}{\delta \tau} + \varphi_{\boldsymbol{x}}{}^{\mathrm{T}}\frac{\delta \boldsymbol{e}_{\boldsymbol{f}}}{\delta \tau} + (\varphi_{\boldsymbol{x}}{}^{\mathrm{T}}\boldsymbol{f}_u + L_{\boldsymbol{u}}{}^{\mathrm{T}})\frac{\delta \boldsymbol{u}}{\delta \tau}\right)\mathrm{d}t\end{aligned} \tag{34}$$

Use Eq. (33), there is

$$\frac{\delta J}{\delta \tau} = (\varphi_t + \varphi_{\boldsymbol{x}}{}^{\mathrm{T}}\dot{\boldsymbol{x}} + L)\Big|_{t_f}\frac{\delta t_f}{\delta \tau} + \int_{t_0}^{t_f}\bar{\boldsymbol{p}}_{\boldsymbol{u}}{}^{\mathrm{T}}\frac{\delta \boldsymbol{u}}{\delta \tau}\mathrm{d}t + \int_{t_0}^{t_f}\left(\boldsymbol{p}_{\boldsymbol{f}}{}^{\mathrm{T}}\frac{\delta \boldsymbol{e}_{\boldsymbol{f}}}{\delta \tau} + \boldsymbol{p}_{\boldsymbol{x}_0}{}^{\mathrm{T}}\frac{\delta \boldsymbol{x}_0}{\delta \tau}\right)\mathrm{d}t \tag{35}$$

where

$$\bar{\boldsymbol{p}}_u = L_{\boldsymbol{u}} + \boldsymbol{f}_u{}^{\mathrm{T}}\varphi_{\boldsymbol{x}} + \boldsymbol{f}_u{}^{\mathrm{T}}\left(\int_t^{t_f}\boldsymbol{\Phi}_o{}^{\mathrm{T}}(\sigma,t)\left(L_{\boldsymbol{x}}(\sigma) + \varphi_{t\boldsymbol{x}}(\sigma) + \varphi_{\boldsymbol{xx}}{}^{\mathrm{T}}(\sigma)\dot{\boldsymbol{x}} + \boldsymbol{f}_{\boldsymbol{x}}(\sigma)^{\mathrm{T}}\varphi_{\boldsymbol{x}}(\sigma)\right)\mathrm{d}\sigma\right) \tag{36}$$

$$\boldsymbol{p}_{\boldsymbol{f}} = \varphi_{\boldsymbol{x}}(t) + \int_t^{t_f}\boldsymbol{\Phi}_o{}^{\mathrm{T}}(\sigma,t)\left(L_{\boldsymbol{x}}(\sigma) + \varphi_{t\boldsymbol{x}}(\sigma) + \varphi_{\boldsymbol{xx}}{}^{\mathrm{T}}(\sigma)\dot{\boldsymbol{x}}(\sigma) + \boldsymbol{f}_{\boldsymbol{x}}(\sigma)^{\mathrm{T}}\varphi_{\boldsymbol{x}}(\sigma)\right)\mathrm{d}\sigma \tag{37}$$

$$\boldsymbol{p}_{\boldsymbol{x}_0} = \boldsymbol{\Phi}_o{}^{\mathrm{T}}(t,t_0)\left(L_{\boldsymbol{x}} + \varphi_{t\boldsymbol{x}} + \varphi_{\boldsymbol{xx}}{}^{\mathrm{T}}\dot{\boldsymbol{x}} + \boldsymbol{f}_{\boldsymbol{x}}{}^{\mathrm{T}}\varphi_{\boldsymbol{x}}\right) \tag{38}$$

In order to be compatible with the derivation in the feasible solution domain, we expect that the searched variation motion of $\frac{\delta \boldsymbol{u}}{\delta \tau}$ and $\frac{\delta t_f}{\delta \tau}$ may negative the first two terms in the right part of Eq. (35).

Now we consider the terminal ECs and start with a simple case that the terminal IECs (17) are always inactive (now $\mathbb{I}_p$ is an empty set and $\boldsymbol{\pi}_I = \boldsymbol{0}$ ), i.e.,

$$\boldsymbol{g}_E\left(\boldsymbol{x}(t_f),t_f\right) \neq \boldsymbol{0} \tag{39}$$

$$\boldsymbol{g}_I\left(\boldsymbol{x}(t_f),t_f\right) < \boldsymbol{0} \tag{40}$$

Then we hope during the evolution, the ECs (39) will satisfy

$$\frac{\delta \boldsymbol{g}_E}{\delta \tau} = (\boldsymbol{g}_E)_{\boldsymbol{x}_f}\frac{\delta \boldsymbol{x}(t_f)}{\delta \tau} + \left((\boldsymbol{g}_E)_{\boldsymbol{x}_f}\dot{\boldsymbol{x}} + (\boldsymbol{g}_E)_{t_f}\right)\frac{\delta t_f}{\delta \tau} = -\boldsymbol{K}_{\boldsymbol{g}_E}\boldsymbol{g}_E \tag{41}$$

where $\boldsymbol{K}_{\boldsymbol{g}_E}$ is a $q_E \times q_E$ dimensional positive-definite matrix and $\frac{\delta \boldsymbol{x}(t_f)}{\delta \tau}$ is calculated by Eq. (33). Follow the thread to get the variation dynamic evolution equation in Ref. [14], we may construct the Feasibility-achieving Evolution Optimization Problem (FAEOP) that realizes Eq. (41) as

**Problem 2**:

$$\begin{aligned}&\min \quad J_{t3} = \frac{1}{2} J_{t1} + \frac{1}{2} J_{t2} \\ &s.t. \\ &\frac{\delta \boldsymbol{g}_E}{\delta \tau} + \boldsymbol{K}_{g_E} \boldsymbol{g}_E = \boldsymbol{0}\end{aligned} \tag{42}$$

where $J_{t1}$ and $J_{t2}$ are defined as

$$J_{t1} = (\varphi_t + \varphi_{\boldsymbol{x}}{}^{\mathrm{T}} \dot{\boldsymbol{x}} + L)\Big|_{t_f} \frac{\delta t_f}{\delta \tau} + \int_{t_0}^{t_f} \bar{\boldsymbol{p}}_{\boldsymbol{u}}{}^{\mathrm{T}} \frac{\delta \boldsymbol{u}}{\delta \tau} \mathrm{d}t \tag{43}$$

$$J_{t2} = \frac{1}{2k_{t_f}} (\frac{\delta t_f}{\delta \tau})^2 + \int_{t_0}^{t_f} \frac{1}{2} (\frac{\delta \boldsymbol{u}}{\delta \tau})^{\mathrm{T}} \boldsymbol{K}^{-1} \frac{\delta \boldsymbol{u}}{\delta \tau} \mathrm{d}t \tag{44}$$

with $\frac{\delta \boldsymbol{u}}{\delta \tau}$ being the optimization variable and $\frac{\delta t_f}{\delta \tau}$ being the optimization parameter.

Solve Problem 2 analytically with the Lagrange multiplier technique, we may derive the evolution equations similar to Eqs. (20) and (21) (with $\boldsymbol{f}$ replaced with $\dot{\boldsymbol{x}}$ and $\boldsymbol{\pi}_I = \boldsymbol{0}$), in which the parameter vector $\boldsymbol{\pi}_E \in \mathbb{R}^{q_E}$ is determined by the linear matrix equation as

$$\boldsymbol{M}\boldsymbol{\pi}_E = -\boldsymbol{r} \tag{45}$$

where the $q_E \times q_E$ dimensional matrix $\boldsymbol{M}$ and the $q_E$ dimensional vector $\boldsymbol{r}$ are

$$\boldsymbol{M} = (\boldsymbol{g}_E)_{\boldsymbol{x}_f} \left( \int_{t_0}^{t_f} \boldsymbol{\Phi}_o(t_f, t) \boldsymbol{f}_{\boldsymbol{u}} \boldsymbol{K} \boldsymbol{f}_{\boldsymbol{u}}{}^{\mathrm{T}} \boldsymbol{\Phi}_o{}^{\mathrm{T}}(t_f, t) \mathrm{d}t \right) (\boldsymbol{g}_E)_{\boldsymbol{x}_f}{}^{\mathrm{T}} + k_{t_f} \left( (\boldsymbol{g}_E)_{\boldsymbol{x}_f} \dot{\boldsymbol{x}} + \boldsymbol{g}_{t_f} \right) \left( (\boldsymbol{g}_E)_{\boldsymbol{x}_f} \dot{\boldsymbol{x}} + \boldsymbol{g}_{t_f} \right)^{\mathrm{T}} \Bigg|_{t_f} \tag{46}$$

$$\begin{aligned}\boldsymbol{r} = {} & (\boldsymbol{g}_E)_{\boldsymbol{x}_f} \left( \int_{t_0}^{t_f} \boldsymbol{\Phi}_o(t_f, t) \boldsymbol{f}_{\boldsymbol{u}} \boldsymbol{K} \boldsymbol{p}_{\boldsymbol{u}} \mathrm{d}t \right) + k_{t_f} \left( (\boldsymbol{g}_E)_{\boldsymbol{x}_f} \dot{\boldsymbol{x}} + (\boldsymbol{g}_E)_{t_f} \right) (\varphi_t + \varphi_{\boldsymbol{x}}{}^{\mathrm{T}} \dot{\boldsymbol{x}} + L)\Big|_{t_f} \\ & - \boldsymbol{K}_{g_E} \boldsymbol{g}_E + (\boldsymbol{g}_E)_{\boldsymbol{x}_f} \boldsymbol{\Phi}_o(t_f, t_0) \boldsymbol{K}_{x_0} \boldsymbol{e}_{x_0} + (\boldsymbol{g}_E)_{\boldsymbol{x}_f} \int_{t_0}^{t_f} \boldsymbol{\Phi}_o(t, s) \boldsymbol{K}_f \boldsymbol{e}_f \mathrm{d}s\end{aligned} \tag{47}$$

In this way, the terminal ECs are expected to be satisfied gradually.

Now we expand the results to further accommodate the violated IECs, that is

$$\boldsymbol{g}_I \left( \boldsymbol{x}(t_f), t_f \right) > \boldsymbol{0} \tag{48}$$

Analogously, we expect that for the violated IECs (and active IECs), their variation motions satisfy

$$\frac{\delta (g_I)_i}{\delta \tau} + (k_{g_I})_i (g_I)_i = \left( \frac{\partial (g_I)_i}{\partial \boldsymbol{x}_f} \right)^{\mathrm{T}} \frac{\delta \boldsymbol{x}(t_f)}{\delta \tau} + \left( \left( \frac{\partial (g_I)_i}{\partial \boldsymbol{x}_f} \right)^{\mathrm{T}} \dot{\boldsymbol{x}} + \frac{\partial (g_I)_i}{\partial t_f} \right) \frac{\delta t_f}{\delta \tau} + (k_{g_I})_i (g_I)_i \leq 0 \quad i \in \mathbb{I} \tag{49}$$

where $\mathbb{I}$ is the index set defined as

$$\mathbb{I} = \{ i \,|\, (g_I)_i \geq 0, \ i = 1, 2, ..., q_I \} \tag{50}$$

and $(k_{g_I})_i$ is a positive constant for the $i$ th terminal IEC. To realize Eq. (49), Problem 2 is adapted as

**Problem 3**:

$$
\begin{aligned}
&\min \quad J_{t3}=\frac{1}{2}J_{t1}+\frac{1}{2}J_{t2} \\
&s.t. \\
&\frac{\delta \boldsymbol{g}_E}{\delta\tau}+\boldsymbol{K}_{\boldsymbol{g}_E}\boldsymbol{g}_E=\boldsymbol{0} \\
&\frac{\delta \boldsymbol{g}_I(\mathbb{I}_p)}{\delta\tau}+\boldsymbol{K}_{\boldsymbol{g}_I}\boldsymbol{g}_I(\mathbb{I}_p)=\boldsymbol{0}
\end{aligned} \tag{51}
$$

where the index set $\mathbb{I}_p$ is modified as

$$
\mathbb{I}_p=\{i\,|\,(g_I)_i\geq 0,\ \frac{\delta(g_I)_i}{\delta\tau}+(k_{g_I})_i(g_I)_i\leq 0 \text{ is an active IEC, } i=1,2,...,q_I\} \tag{52}
$$

$\boldsymbol{K}_{\boldsymbol{g}_I}=\mathrm{diag}\left(\boldsymbol{k}_{\boldsymbol{g}_I}(\mathbb{I}_p)\right)$ is a $n_{\mathbb{I}_p}\times n_{\mathbb{I}_p}$ dimensional positive-definite diagonal matrix and $\boldsymbol{k}_{\boldsymbol{g}_I}=\left[(k_{g_I})_1 \quad (k_{g_I})_2 \quad ... \quad (k_{g_I})_{q_I}\right]^{\mathrm{T}}$.

From Problem 3, we hope to obtain the evolution equations that may eliminate the violation on the terminal IECs. However, for this adapted FAEOP, one may argue that

i) Since the IECs with index in $\mathbb{I}$ are all violated, why only components in $\mathbb{I}_P$ are considered in Problem 3?

ii) Using the infinite-time asymptotically convergence evolution, the violated IECs may never return to the feasible domain. Thus a limited-time convergence dynamics should be used in Problem 3 as

$$
\frac{\delta \boldsymbol{g}_I(\mathbb{I}_p)}{\delta\tau}+\boldsymbol{K}_{\boldsymbol{g}_I}\mathrm{sign}\left(\boldsymbol{g}_I(\mathbb{I}_p)\right)=\boldsymbol{0} \tag{53}
$$

where $\mathrm{sign}(\cdot)$ is the sign function.

For these argument, we will show that even if an infinite-time convergence dynamics is employed for the IECs in $\mathbb{I}_P$, all the violated IECs in $\mathbb{I}$ will succeed in approaching their optimal value. Moreover, although Eq. (53) is usable theoretically, it may result in the unexpected chattering that is disadvantageous to the numerical computation.

**Proposition 1**: The evolution equations derived from Problem 3 guarantee that the terminal IECs with index in $\hat{\mathbb{I}}_p$ (See Sec. III.B for its definition) will be achieved asymptotically, and the IECs not in $\hat{\mathbb{I}}_p$ will return to the feasible domain in finite time.

**Proof**: For an active IEC of the FAEOP, it means that

$$
\frac{\delta(g_I)_i}{\delta\tau}+(k_{g_I})_i(g_I)_i=0 \quad i\in\mathbb{I}_p \tag{54}
$$

Under such dynamics, $(g_I)_i$ will gradually approach zero. Since $\mathbb{I}_p$ will evolve to $\hat{\mathbb{I}}_p$ ultimately, this means that the IECs with index in $\hat{\mathbb{I}}_p$ will be achieved asymptotically.

For an inactive IEC of the FAEOP, we have (See Ref. [15])

$$
\frac{\delta(g_I)_i}{\delta\tau}<-(k_{g_I})_i(g_I)_i \quad i\notin\mathbb{I}_p \tag{55}
$$

On the other hand, for an IEC $(g_I)_i$, which will enter the inactive domain from $(g_I)_i=0$, we have

$$
\frac{\delta(g_I)_i}{\delta\tau}<-a \quad when \ (g_I)_i=0 \tag{56}
$$

where $a$ is some positive constant. Thus, there exists a neighborhood $[0,\varepsilon]$ such that

$$\frac{\delta(g_I)_i}{\delta\tau} < -\frac{a}{2} \quad when \ (g_I)_i \in [0,\varepsilon] \tag{57}$$

where $\varepsilon$ is some positive constant. Under the dynamics of Eq. (55), $(g_I)_i$ will enter $[0,\varepsilon]$ in finite time from any positive value. After that, $(g_I)_i$ will reach zero with time smaller than $\frac{2\varepsilon}{a}$ from $(g_I)_i = \varepsilon$. Thus, for those IECs not in $\hat{\mathbb{I}}_p$, they will return to the feasible domain in a limited time. ■

Through solving Problem 3, we may derive the modified evolution equations as

$$\frac{\delta \boldsymbol{u}}{\delta\tau} = -\boldsymbol{K}\overline{\boldsymbol{p}}_{\boldsymbol{u}}^{tc} \tag{58}$$

$$\frac{\delta t_f}{\delta\tau} = -k_{t_f}\, \overline{p}_{t_f}^{tc} \tag{59}$$

where

$$\overline{\boldsymbol{p}}_{\boldsymbol{u}}^{tc} = \overline{\boldsymbol{p}}_{\boldsymbol{u}} + \boldsymbol{f_u}^{\mathrm{T}}\boldsymbol{\Phi}_o^{\mathrm{T}}(t_f,t)\Big((\boldsymbol{g}_E)_{\boldsymbol{x}_f}{}^{\mathrm{T}}\boldsymbol{\pi}_E + (\boldsymbol{g}_I)_{\boldsymbol{x}_f}{}^{\mathrm{T}}\boldsymbol{\pi}_I\Big) \tag{60}$$

$$\overline{p}_{t_f}^{tc} = \Big(L + \varphi_t + \varphi_{\boldsymbol{x}}{}^{\mathrm{T}}\dot{\boldsymbol{x}} + \boldsymbol{\pi}_E{}^{\mathrm{T}}\Big((\boldsymbol{g}_E)_{\boldsymbol{x}_f}\dot{\boldsymbol{x}} + (\boldsymbol{g}_E)_{t_f}\Big) + \boldsymbol{\pi}_I{}^{\mathrm{T}}\Big((\boldsymbol{g}_I)_{\boldsymbol{x}_f}\dot{\boldsymbol{x}} + (\boldsymbol{g}_I)_{t_f}\Big)\Big)\Big|_{t_f} \tag{61}$$

The parameters $\boldsymbol{\pi}_E$ and $\boldsymbol{\pi}_I$ are also determined by Eq. (25), with the index set $\mathbb{I}_p$ defined in Eq. (52), and Eqs. (28), (29) correspondingly modified as

$$\boldsymbol{M} = \boldsymbol{g}_{\boldsymbol{x}_f}\left(\int_{t_0}^{t_f}\boldsymbol{\Phi}_o(t_f,t)\boldsymbol{f_u}\boldsymbol{K}\boldsymbol{f_u}^{\mathrm{T}}\boldsymbol{\Phi}_o^{\mathrm{T}}(t_f,t)\,\mathrm{d}t\right)\boldsymbol{g}_{\boldsymbol{x}_f}{}^{\mathrm{T}} + k_{t_f}(\boldsymbol{g}_{\boldsymbol{x}_f}\dot{\boldsymbol{x}} + \boldsymbol{g}_{t_f})(\boldsymbol{g}_{\boldsymbol{x}_f}\dot{\boldsymbol{x}} + \boldsymbol{g}_{t_f})^{\mathrm{T}}\Big|_{t_f} \tag{62}$$

$$\begin{aligned}\boldsymbol{r} = {} & \boldsymbol{g}_{\boldsymbol{x}_f}\left(\int_{t_0}^{t_f}\boldsymbol{\Phi}_o(t_f,t)\boldsymbol{f_u}\boldsymbol{K}\boldsymbol{p_u}\,\mathrm{d}t\right) + k_{t_f}(\boldsymbol{g}_{\boldsymbol{x}_f}\dot{\boldsymbol{x}} + \boldsymbol{g}_{t_f})(\varphi_t + \varphi_{\boldsymbol{x}}{}^{\mathrm{T}}\dot{\boldsymbol{x}} + L)\Big|_{t_f} \\ & - \begin{bmatrix}\boldsymbol{K}_{g_E}\boldsymbol{g}_E \\ \boldsymbol{K}_{g_I}\boldsymbol{g}_I(\mathbb{I}_p)\end{bmatrix} + \boldsymbol{g}_{\boldsymbol{x}_f}\boldsymbol{\Phi}_o(t_f,t_0)\boldsymbol{K}_{x_0}\boldsymbol{e}_{x_0} + \boldsymbol{g}_{\boldsymbol{x}_f}\int_{t_0}^{t_f}\boldsymbol{\Phi}_o(t,s)\boldsymbol{K}_f\boldsymbol{e}_{\boldsymbol{f}}\,\mathrm{d}s\end{aligned} \tag{63}$$

Note that $\boldsymbol{g} = \begin{bmatrix}\boldsymbol{g}_E \\ \boldsymbol{g}_I(\mathbb{I}_p)\end{bmatrix}$.

We again return to the change of the performance index (13) in the infeasible solution domain. Now Eq. (35) may be re-presented as

$$\frac{\delta J}{\delta\tau} = \overline{p}_{t_f}^{tc}\frac{\delta t_f}{\delta\tau} + \int_{t_0}^{t_f}(\overline{\boldsymbol{p}}_{\boldsymbol{u}}^{tc})^{\mathrm{T}}\frac{\delta\boldsymbol{u}}{\delta\tau}\mathrm{d}t + \int_{t_0}^{t_f}\left(\boldsymbol{p_f}^{\mathrm{T}}\frac{\delta\boldsymbol{e_f}}{\delta\tau} + \boldsymbol{p}_{x_0}{}^{\mathrm{T}}\frac{\delta\boldsymbol{x}_0}{\delta\tau}\right)\mathrm{d}t - \boldsymbol{\pi}_E{}^{\mathrm{T}}\frac{\delta\boldsymbol{g}_E}{\delta\tau} - \boldsymbol{\pi}_I{}^{\mathrm{T}}\frac{\delta\boldsymbol{g}_I}{\delta\tau} \tag{64}$$

By investigating Eq. (64), it is found that starting from an infeasible solution, $J$ will not monotonously decrease even under Eqs. (58) and (59), because the terms $\int_{t_0}^{t_f}\left(\boldsymbol{p_f}^{\mathrm{T}}\frac{\delta\boldsymbol{e_f}}{\delta\tau} + \boldsymbol{p}_{x_0}{}^{\mathrm{T}}\frac{\delta\boldsymbol{x}_0}{\delta\tau}\right)\mathrm{d}t - \boldsymbol{\pi}_E{}^{\mathrm{T}}\frac{\delta\boldsymbol{g}_E}{\delta\tau} - \boldsymbol{\pi}_I{}^{\mathrm{T}}\frac{\delta\boldsymbol{g}_I}{\delta\tau}$ arising from infeasibilities may be positive, and the sign of $\frac{\delta J}{\delta\tau}$ is uncertain.

*D. Mathematic validation*

With the modification in last sub-section, we anticipate that the modified variation dynamic evolution equations (33), (58) and (59) will evolve an arbitrary initial guess of solutions to the optimal, by achieving the feasibility and optimality simultaneously. It is not hard to verify that optimal solutions satisfying the feasibility conditions (14)-(17) and optimality conditions (30), (31) are the equilibrium solutions of Eqs. (33), (58) and (59). However, lacking the convergence guarantee by Lemma 1 (with Eq. (13) as the Lyapunov functional), it is natural to ask that under the dynamics governed by Eqs. (33), (58) and (59), is it ensured that the variables will approach the equilibrium solution from arbitrary initial value, instead of converging to the limit cycle as the Van der Pol oscillator [18]? Now we will answer this question with rigorous mathematic argument as follows. Before we carry out the mathematic analysis, certain assumptions are presented.

**Assumption 1**: The solutions during the variable evolution satisfy the controllability requirement [22], i.e., the existence of the solution for Eq. (27) is guaranteed.

**Assumption 2**: During the evolution process, the variables $\boldsymbol{p}_f(t)$ and $\boldsymbol{p}_{\boldsymbol{x}_0}(t)$, defined in Eqs. (37) and (38) respectively, are bounded within the time horizon $[t_0, t_f]$ as

$$\left\| \boldsymbol{p}_f(t) \right\|_2 \le d_1 \qquad t \in [t_0, t_f] \tag{65}$$

$$\left\| \boldsymbol{p}_{\boldsymbol{x}_0}(t) \right\|_2 \le d_2 \qquad t \in [t_0, t_f] \tag{66}$$

The multiplier parameters $\boldsymbol{\pi}_E$ and $\boldsymbol{\pi}_I$ determined in Eq. (25) are bounded as

$$\left\| \boldsymbol{\pi}_E \right\|_2 \le d_3 \tag{67}$$

$$\left\| \boldsymbol{\pi}_I \right\|_2 \le d_4 \tag{68}$$

where $\left\| \cdot \right\|_2$ denotes the 2-norm of vector.

**Lemma 2**: For the unconstrained functional

$$V = \sqrt{\boldsymbol{e}_{\boldsymbol{x}_0}{}^{\mathrm{T}} \boldsymbol{e}_{\boldsymbol{x}_0}} + \int_{t_0}^{t_f} \sqrt{\boldsymbol{e}_f{}^{\mathrm{T}} \boldsymbol{e}_f} \mathrm{d}t + \sqrt{\boldsymbol{g}_E{}^{\mathrm{T}} \boldsymbol{g}_E} + \sqrt{\boldsymbol{g}_I(\mathbb{I})^{\mathrm{T}} \boldsymbol{g}_I(\mathbb{I})} + c_1 J + \frac{1}{2} c_2 \left( \boldsymbol{e}_f{}^{\mathrm{T}} \boldsymbol{e}_f \right)\Big|_{t_f} \tag{69}$$

where $\boldsymbol{e}_{\boldsymbol{x}_0} = \boldsymbol{x}(t_0) - \boldsymbol{x}_0$, $\boldsymbol{e}_f = \dot{\boldsymbol{x}} - \boldsymbol{f}(\boldsymbol{x}, \boldsymbol{u}, t)$, $J$ is defined in Eq. (13) and $\mathbb{I}$ is defined in Eq. (50), there exists certain positive constants $c_1$ and $c_2$

$$c_1 < \min\left( \frac{\min(\mathrm{eig}(\boldsymbol{K}_f))}{d_1 \max(\mathrm{eig}(\boldsymbol{K}_f))(t_f - t_0)}, \frac{\min(\mathrm{eig}(\boldsymbol{K}_{\boldsymbol{x}_0}))}{d_2 \max(\mathrm{eig}(\boldsymbol{K}_{\boldsymbol{x}_0}))(t_f - t_0)^2}, \frac{\min(\mathrm{eig}(\boldsymbol{K}_{\boldsymbol{g}_E}))}{d_3 \max(\mathrm{eig}(\boldsymbol{K}_{\boldsymbol{g}_E}))}, \frac{\min(\boldsymbol{k}_{\boldsymbol{g}_I})}{d_4 \max(\mathrm{eig}(\boldsymbol{K}_{\boldsymbol{g}_I}))} \right) \tag{70}$$

$$c_2 > \frac{k_{t_f}}{2 c_1 \min(\mathrm{eig}(\boldsymbol{K}_f))} \tag{71}$$

such that Eq.(69) is a Lyapunov functional for the variation dynamic evolution equations (33), (58) and (59). Here $\mathrm{eig}(\cdot)$ is the function of eigenvalue.

**Proof**: First we show that the minimum solution of Problem 1, denoted by $(\hat{\boldsymbol{x}}, \hat{\boldsymbol{u}})$, is also the minimum solution of the unconstrained functional (69). If the variables $\boldsymbol{x}$ and $\boldsymbol{u}$ are located within the feasible domain, in which the variables meet Eqs. (14)-(17), then we have

$$V = c_1 J \tag{72}$$

Obviously for this case the minimum of Problem 1 is the minimum of the unconstrained functional (69). When the variables lie in the infeasible domain, we consider the neighborhood around the minimum solution $(\hat{\boldsymbol{x}}, \hat{\boldsymbol{u}})$. Since $(\hat{\boldsymbol{x}}, \hat{\boldsymbol{u}})$ satisfies Eqs. (14)-(17), (30), and (31), we have the first order variation of the functional (69) at $(\hat{\boldsymbol{x}}, \hat{\boldsymbol{u}})$ as

$$\delta V = \left\|\delta \boldsymbol{e}_{\boldsymbol{x}_0}\right\|_2 + \int_{t_0}^{t_f} \left\|\delta \boldsymbol{e}_{\boldsymbol{f}}\right\|_2 \mathrm{d}t + \left\|\delta \boldsymbol{g}_E\right\|_2 + \left\|\delta \boldsymbol{g}_I(\mathbb{I})\right\|_2 + c_1 \int_{t_0}^{t_f} \boldsymbol{p}_{\boldsymbol{f}}^{\mathrm{T}} \delta \boldsymbol{e}_{\boldsymbol{f}} \mathrm{d}t + c_1 \int_{t_0}^{t_f} \boldsymbol{p}_{\boldsymbol{x}_0}^{\mathrm{T}} \delta \boldsymbol{e}_{\boldsymbol{x}_0} \mathrm{d}t - c_1 \boldsymbol{\pi}_E^{\mathrm{T}} \delta \boldsymbol{g}_E - c_1 \boldsymbol{\pi}_I(\mathbb{I}_p)^{\mathrm{T}} \delta \boldsymbol{g}_I(\mathbb{I}_p) \tag{73}$$

Note that

$$\boldsymbol{\pi}_I^{\mathrm{T}} \delta \boldsymbol{g}_I = \boldsymbol{\pi}_I(\mathbb{I}_p)^{\mathrm{T}} \delta \boldsymbol{g}_I(\mathbb{I}_p) \tag{74}$$

According to Assumptions 2, and with the Holder's inequality, there are

$$-d_1(t_f - t_0) \int_{t_0}^{t_f} \left\|\delta \boldsymbol{e}_{\boldsymbol{f}}\right\|_2 \mathrm{d}t \le -\int_{t_0}^{t_f} \left\|\boldsymbol{p}_{\boldsymbol{f}}\right\|_2 \mathrm{d}t \int_{t_0}^{t_f} \left\|\delta \boldsymbol{e}_{\boldsymbol{f}}\right\|_2 \mathrm{d}t \le \int_{t_0}^{t_f} \boldsymbol{p}_{\boldsymbol{f}}^{\mathrm{T}} \delta \boldsymbol{e}_{\boldsymbol{f}} \mathrm{d}t \tag{75}$$

$$-d_2(t_f - t_0)^2 \left\|\delta \boldsymbol{e}_{\boldsymbol{x}_0}\right\|_2 \le -\int_{t_0}^{t_f} \left\|\boldsymbol{p}_{\boldsymbol{x}_0}\right\|_2 \mathrm{d}t \int_{t_0}^{t_f} \left\|\delta \boldsymbol{e}_{\boldsymbol{x}_0}\right\|_2 \mathrm{d}t \le \int_{t_0}^{t_f} \boldsymbol{p}_{\boldsymbol{x}_0}^{\mathrm{T}} \delta \boldsymbol{e}_{\boldsymbol{x}_0} \mathrm{d}t \tag{76}$$

$$-d_3 \left\|\delta \boldsymbol{g}_E\right\|_2 \le -\left\|\boldsymbol{\pi}_E\right\|_2 \left\|\delta \boldsymbol{g}_E\right\|_2 \le -\boldsymbol{\pi}_E^{\mathrm{T}} \delta \boldsymbol{g}_E \tag{77}$$

$$-d_4 \left\|\delta \boldsymbol{g}_I(\mathbb{I})\right\|_2 \le -\left\|\boldsymbol{\pi}_I\right\|_2 \left\|\delta \boldsymbol{g}_I(\mathbb{I}_p)\right\|_2 \le -\boldsymbol{\pi}_I(\mathbb{I}_p)^{\mathrm{T}} \delta \boldsymbol{g}_I(\mathbb{I}_p) \tag{78}$$

Then we have

$$\delta V \ge \left(1 - c_1 d_2 (t_f - t_0)^2\right) \left\|\delta \boldsymbol{e}_{\boldsymbol{x}_0}\right\|_2 + \left(1 - c_1 d_1 (t_f - t_0)\right) \int_{t_0}^{t_f} \left\|\delta \boldsymbol{e}_{\boldsymbol{f}}\right\|_2 \mathrm{d}t + (1 - c_1 d_3) \left\|\delta \boldsymbol{g}_E\right\|_2 + (1 - c_1 d_4) \left\|\delta \boldsymbol{g}_I(\mathbb{I})\right\|_2 \tag{79}$$

According to Eq. (70), we have $\delta V > 0$. Especially, since $c_1$ may be arbitrarily small, the infeasible domain where the minimum maintains may be arbitrarily large. In summary, the solution $(\hat{\boldsymbol{x}}, \hat{\boldsymbol{u}})$ determines a minimum for the functional (69).

Now we consider the derivative of $V$ with respect to the variation time $\tau$. Differentiating Eq. (69) produces

$$\begin{aligned} \frac{\delta V}{\delta \tau} = & \frac{\boldsymbol{e}_{\boldsymbol{x}_0}^{\mathrm{T}}}{\sqrt{\boldsymbol{e}_{\boldsymbol{x}_0}^{\mathrm{T}} \boldsymbol{e}_{\boldsymbol{x}_0}}} \frac{\delta \boldsymbol{x}(t_0)}{\delta \tau} + \int_{t_0}^{t_f} \frac{\boldsymbol{e}_{\boldsymbol{f}}^{\mathrm{T}}}{\sqrt{\boldsymbol{e}_{\boldsymbol{f}}^{\mathrm{T}} \boldsymbol{e}_{\boldsymbol{f}}}} \frac{\delta \boldsymbol{e}_{\boldsymbol{f}}}{\delta \tau} \mathrm{d}t + \frac{\boldsymbol{g}_E^{\mathrm{T}}}{\sqrt{\boldsymbol{g}_E^{\mathrm{T}} \boldsymbol{g}_E}} \frac{\delta \boldsymbol{g}_E}{\delta \tau} + \frac{\boldsymbol{g}_I(\mathbb{I})^{\mathrm{T}}}{\sqrt{\boldsymbol{g}_I(\mathbb{I})^{\mathrm{T}} \boldsymbol{g}_I(\mathbb{I})}} \frac{\delta \boldsymbol{g}_I(\mathbb{I})}{\delta \tau} \\ & + c_1 \frac{\delta J}{\delta \tau} + c_2 \left(\boldsymbol{e}_{\boldsymbol{f}}^{\mathrm{T}} \frac{\delta \boldsymbol{e}_{\boldsymbol{f}}}{\delta \tau}\right)\Big|_{t_f} + \left(\sqrt{\boldsymbol{e}_{\boldsymbol{f}}^{\mathrm{T}} \boldsymbol{e}_{\boldsymbol{f}}}\right)\Big|_{t_f} \frac{\delta t_f}{\delta \tau} \end{aligned} \tag{80}$$

Substitute Eq. (64) in and use Eqs. (33), (58) and (59), and especially note that

$$\boldsymbol{g}_I(\mathbb{I})^{\mathrm{T}} \frac{\delta \boldsymbol{g}_I(\mathbb{I})}{\delta \tau} \le \boldsymbol{g}_I(\mathbb{I})^{\mathrm{T}} \mathrm{diag}\left(\boldsymbol{k}_{g_I}(\mathbb{I})\right) \boldsymbol{g}_I(\mathbb{I}) \tag{81}$$

$$\boldsymbol{\pi}_I^{\mathrm{T}} \frac{\delta \boldsymbol{g}_I}{\delta \tau} = \boldsymbol{\pi}_I(\mathbb{I}_p)^{\mathrm{T}} \boldsymbol{K}_{g_I} \boldsymbol{g}_I(\mathbb{I}_p) \tag{82}$$

Then we have

$$\begin{aligned} \frac{\delta V}{\delta \tau} \le & -\frac{\boldsymbol{e}_{\boldsymbol{x}_0}^{\mathrm{T}}}{\sqrt{\boldsymbol{e}_{\boldsymbol{x}_0}^{\mathrm{T}} \boldsymbol{e}_{\boldsymbol{x}_0}}} \boldsymbol{K}_{x_0} \boldsymbol{e}_{x_0} - \int_{t_0}^{t_f} \frac{\boldsymbol{e}_{\boldsymbol{f}}^{\mathrm{T}}}{\sqrt{\boldsymbol{e}_{\boldsymbol{f}}^{\mathrm{T}} \boldsymbol{e}_{\boldsymbol{f}}}} \boldsymbol{K}_f \boldsymbol{e}_{\boldsymbol{f}} \mathrm{d}t - \frac{\boldsymbol{g}_E^{\mathrm{T}}}{\sqrt{\boldsymbol{g}_E^{\mathrm{T}} \boldsymbol{g}_E}} \boldsymbol{K}_{g_E} \boldsymbol{g}_E - \frac{\boldsymbol{g}_I(\mathbb{I})^{\mathrm{T}}}{\sqrt{\boldsymbol{g}_I(\mathbb{I})^{\mathrm{T}} \boldsymbol{g}_I(\mathbb{I})}} \mathrm{diag}\left(\boldsymbol{k}_{g_I}(\mathbb{I})\right) \boldsymbol{g}_I(\mathbb{I}) - k_{t_f} c_1 (\bar{p}_{t_f}^{tc})^2 - c_1 \int_{t_0}^{t_f} (\bar{\boldsymbol{p}}_{\boldsymbol{u}}^{tc})^{\mathrm{T}} \boldsymbol{K} \bar{\boldsymbol{p}}_{\boldsymbol{u}}^{tc} \mathrm{d}t \\ & - c_1 \int_{t_0}^{t_f} \boldsymbol{p}_{\boldsymbol{f}}^{\mathrm{T}} \boldsymbol{K}_f \boldsymbol{e}_{\boldsymbol{f}} \mathrm{d}t - c_1 \int_{t_0}^{t_f} \boldsymbol{p}_{x_0}^{\mathrm{T}} \boldsymbol{K}_{x_0} \boldsymbol{e}_{\boldsymbol{x}_0} \mathrm{d}t + c_1 \boldsymbol{\pi}_E^{\mathrm{T}} \boldsymbol{K}_{g_E} \boldsymbol{g}_E + c_1 \boldsymbol{\pi}_I(\mathbb{I}_p)^{\mathrm{T}} \boldsymbol{K}_{g_I} \boldsymbol{g}_I(\mathbb{I}_p) - c_2 \left(\boldsymbol{e}_{\boldsymbol{f}}^{\mathrm{T}} \boldsymbol{K}_f \boldsymbol{e}_{\boldsymbol{f}}\right)\Big|_{t_f} - k_{t_f} \bar{p}_{t_f}^{tc} \sqrt{\boldsymbol{e}_{\boldsymbol{f}}^{\mathrm{T}} \boldsymbol{e}_{\boldsymbol{f}}} \end{aligned} \tag{83}$$

With the Young's inequality, there is

$$-k_{t_f}\bar{p}_{t_f}^{tc}\sqrt{\boldsymbol{e_f}^{\mathrm{T}}\boldsymbol{e_f}} \le \frac{k_{t_f}c_1}{2}(\bar{p}_{t_f}^{tc})^2 + \frac{k_{t_f}}{2c_1}(\left\|\boldsymbol{e_f}\right\|_2^2)\Big|_{t_f} \tag{84}$$

According to Assumptions 2, and with the Holder's inequality, there are

$$-\int_{t_0}^{t_f}\boldsymbol{p_f}^{\mathrm{T}}\boldsymbol{K}_f\boldsymbol{e_f}\mathrm{d}t \le \max(\mathrm{eig}(\boldsymbol{K}_f))\left(\int_{t_0}^{t_f}\left\|\boldsymbol{p_f}\right\|_2\mathrm{d}t\right)\int_{t_0}^{t_f}\left\|\boldsymbol{e_f}\right\|_2\mathrm{d}t \le d_1\max(\mathrm{eig}(\boldsymbol{K}_f))(t_f-t_0)\int_{t_0}^{t_f}\left\|\boldsymbol{e_f}\right\|_2\mathrm{d}t \tag{85}$$

$$-\int_{t_0}^{t_f}\boldsymbol{p}_{\boldsymbol{x}_0}^{\mathrm{T}}\boldsymbol{K}_{x_0}\boldsymbol{e}_{\boldsymbol{x}_0}\mathrm{d}t \le \max(\mathrm{eig}(\boldsymbol{K}_{x_0}))(t_f-t_0)\left(\int_{t_0}^{t_f}\left\|\boldsymbol{p}_{\boldsymbol{x}_0}\right\|_2\mathrm{d}t\right)\left\|\boldsymbol{e}_{\boldsymbol{x}_0}\right\|_2 \le d_2\max(\mathrm{eig}(\boldsymbol{K}_{x_0}))(t_f-t_0)^2\left\|\boldsymbol{e}_{\boldsymbol{x}_0}\right\|_2 \tag{86}$$

$$\boldsymbol{\pi}_E^{\mathrm{T}}\boldsymbol{K}_{\boldsymbol{g}_E}\boldsymbol{g}_E \le \max(\mathrm{eig}(\boldsymbol{K}_{\boldsymbol{g}_E}))\left\|\boldsymbol{\pi}_E\right\|_2\left\|\boldsymbol{g}_E\right\|_2 \le d_3\max(\mathrm{eig}(\boldsymbol{K}_{\boldsymbol{g}_E}))\left\|\boldsymbol{g}_E\right\|_2 \tag{87}$$

In particular, there is

$$\boldsymbol{\pi}_I(\mathbb{I}_p)^{\mathrm{T}}\boldsymbol{K}_{\boldsymbol{g}_I}\boldsymbol{g}_I(\mathbb{I}_p) \le \max(\mathrm{eig}(\boldsymbol{K}_{\boldsymbol{g}_I}))\left\|\boldsymbol{\pi}_I\right\|_2\left\|\boldsymbol{g}_I(\mathbb{I}_p)\right\|_2 \le d_4\max(\mathrm{eig}(\boldsymbol{K}_{\boldsymbol{g}_I}))\left\|\boldsymbol{g}_I(\mathbb{I})\right\|_2 \tag{88}$$

Substituting the inequalities (84)-(88) into Eq. (83) gives

$$\begin{aligned}\frac{\delta V}{\delta\tau} \le &-\left(\min(\mathrm{eig}(\boldsymbol{K}_{x_0}))-c_1d_2\max(\mathrm{eig}(\boldsymbol{K}_{x_0}))(t_f-t_0)^2\right)\left\|\boldsymbol{e}_{\boldsymbol{x}_0}\right\|_2-\left(\min(\mathrm{eig}(\boldsymbol{K}_f))-c_1d_1\max(\mathrm{eig}(\boldsymbol{K}_f))(t_f-t_0)\right)\int_{t_0}^{t_f}\left\|\boldsymbol{e_f}\right\|_2\mathrm{d}t\\ &-\left(\min(\mathrm{eig}(\boldsymbol{K}_{\boldsymbol{g}_E}))-c_1d_3\max(\mathrm{eig}(\boldsymbol{K}_{\boldsymbol{g}_E}))\right)\left\|\boldsymbol{g}_E\right\|_2-\left(\min(\boldsymbol{k}_{\boldsymbol{g}_I})-c_1d_4\max(\mathrm{eig}(\boldsymbol{K}_{\boldsymbol{g}_I}))\right)\left\|\boldsymbol{g}_I(\mathbb{I})\right\|_2\\ &-\left(c_2\min(\mathrm{eig}(\boldsymbol{K}_f))-\frac{k_{t_f}}{2c_1}\right)(\left\|\boldsymbol{e_f}\right\|_2^2)\Big|_{t_f}-c_1\int_{t_0}^{t_f}(\bar{\boldsymbol{p}}_{\boldsymbol{u}}^{tc})^{\mathrm{T}}\boldsymbol{K}\bar{\boldsymbol{p}}_{\boldsymbol{u}}^{tc}\mathrm{d}t-\frac{k_{t_f}c_1}{2}(\bar{p}_{t_f}^{tc})^2\end{aligned} \tag{89}$$

With the values of $c_1$ and $c_2$ set by Eqs. (70) and (71), we have that $\frac{\delta V}{\delta\tau}\le 0$ hold under the dynamics governed by Eqs. (33), (58) and (59), and $\frac{\delta V}{\delta\tau}=0$ when Eqs. (14)-(17), (30), and (31) are satisfied. ■

**Theorem 1:** Solving the IVP with respect to $\tau$, defined by the variation dynamic evolution equations (33), (58) and (59) with any initial solution, when $\tau\to+\infty$, $(\boldsymbol{x},\boldsymbol{u})$ will satisfy the feasibility conditions and the optimality conditions of Problem 1.

**Proof**: The proof is a direct application of Lemmas 1 and 2. From Lemma 2, the functional (69) is ensured a Lyapunov functional for the dynamic system (33), (58) and (59) around the equilibrium that meets (14)-(17), (30), and (31). According to Lemma 1, the equilibrium solutions is an asymptotically stable solution and $(\boldsymbol{x},\boldsymbol{u})$ will satisfy the feasibility conditions (14)-(17), and the optimality conditions (30), (31) of Problem 1 when $\tau\to+\infty$. ■

*E. Formulation of MEPDE*

Use the partial differential operator "$\partial$" and the differential operator "$\mathrm{d}$" to reformulate the variation dynamic evolution equations, we may get the MEPDEs including (9) and

$$\frac{\partial\boldsymbol{u}}{\partial\tau}=-\boldsymbol{K}\left\{\begin{aligned}&L_{\boldsymbol{u}}+\boldsymbol{f}_u^{\mathrm{T}}\varphi_{\boldsymbol{x}}+\boldsymbol{f_u}^{\mathrm{T}}\boldsymbol{\Phi}_o^{\mathrm{T}}(t_f,t)\left((\boldsymbol{g}_E)_{x_f}^{\mathrm{T}}\boldsymbol{\pi}_E+(\boldsymbol{g}_I)_{x_f}^{\mathrm{T}}\boldsymbol{\pi}_I\right)\\ &+\boldsymbol{f}_u^{\mathrm{T}}\int_t^{t_f}\boldsymbol{\Phi}_o^{\mathrm{T}}(\sigma,t)\left(L_{\boldsymbol{x}}(\sigma)+\varphi_{t\boldsymbol{x}}(\sigma)+\varphi_{\boldsymbol{xx}}^{\mathrm{T}}(\sigma)\frac{\partial\boldsymbol{x}}{\partial t}(\sigma)+\boldsymbol{f}_{\boldsymbol{x}}(\sigma)^{\mathrm{T}}\varphi_{\boldsymbol{x}}(\sigma)\right)\mathrm{d}\sigma\end{aligned}\right\} \tag{90}$$

and the Modified EDE (MEDE) as

$$\frac{\mathrm{d}t_f}{\mathrm{d}\tau}=-k_{t_f}\left(L+\varphi_t+\varphi_{\boldsymbol{x}}^{\mathrm{T}}\frac{\partial\boldsymbol{x}}{\partial t}+\boldsymbol{\pi}_E^{\mathrm{T}}\left((\boldsymbol{g}_E)_{x_f}\frac{\partial\boldsymbol{x}}{\partial t}+(\boldsymbol{g}_E)_{t_f}\right)+\boldsymbol{\pi}_I^{\mathrm{T}}\left((\boldsymbol{g}_I)_{x_f}\frac{\partial\boldsymbol{x}}{\partial t}+(\boldsymbol{g}_I)_{t_f}\right)\right)\Bigg|_{t_f} \tag{91}$$

In particular, the definite conditions $\begin{bmatrix}\boldsymbol{x}(t,\tau)\\ \boldsymbol{u}(t,\tau)\end{bmatrix}\Bigg|_{\tau=0}$ and $t_f\big|_{\tau=0}$ may be arbitrary solutions.

Recall the anticipated variable evolution along the variation time $\tau$ illustrated in Fig. 1, the initial conditions of $\boldsymbol{x}(t,\tau)$ and $\boldsymbol{u}(t,\tau)$ at $\tau=0$ may be infeasible and their value at $\tau=+\infty$ will be the optimal solution of the OCP. The right part of the MEPDEs are also only vector functions of time $t$. Thus we may apply the semi-discrete method to discretize them along the normal time dimension and further use ODE integration methods to get the numerical solution. Moreover, the results obtained in this paper are also applicable to the OCPs with fixed terminal time. By setting $k_{t_f}=0$, these equations may be directly applied.

## IV. Illustrative Examples

First a linear example taken from Xie [23] is considered.

**Example 1**: Consider the following dynamic system

$$\dot{\boldsymbol{x}}=\boldsymbol{A}\boldsymbol{x}+\boldsymbol{b}u$$

where $\boldsymbol{x}=\begin{bmatrix}x_1\\x_2\end{bmatrix}$, $\boldsymbol{A}=\begin{bmatrix}0&1\\0&0\end{bmatrix}$, and $\boldsymbol{b}=\begin{bmatrix}0\\1\end{bmatrix}$. Find the solution that minimizes the performance index

$$J=\frac{1}{2}\int_{t_0}^{t_f}u^2\mathrm{d}t$$

with the boundary conditions

$$\boldsymbol{x}(t_0)=\begin{bmatrix}1\\1\end{bmatrix},\ \boldsymbol{x}(t_f)=\begin{bmatrix}0\\0\end{bmatrix}$$

where the initial time $t_0=0$ and the terminal time $t_f=2$ are fixed.

In solving this example using the VEM, the MEPDEs derived are

$$\frac{\partial}{\partial\tau}\begin{bmatrix}\boldsymbol{x}\\u\end{bmatrix}=\begin{bmatrix}-e^{\boldsymbol{A}(t-t_0)}\boldsymbol{K}_{x_0}\left(\boldsymbol{x}(t_0)-\boldsymbol{x}_0\right)+\int_{t_0}^{t}e^{\boldsymbol{A}(t-s)}\boldsymbol{b}\frac{\partial\boldsymbol{u}}{\partial\tau}(s)\,\mathrm{d}s-\int_{t_0}^{t}e^{\boldsymbol{A}(t-s)}\boldsymbol{K}_f\left(\frac{\partial\boldsymbol{x}}{\partial t}(s)-\boldsymbol{A}\boldsymbol{x}(s)-\boldsymbol{b}u(s)\right)\mathrm{d}s\\-K\left\{u+\boldsymbol{b}^{\mathrm{T}}\left(e^{\boldsymbol{A}(t_f-t)}\right)^{\mathrm{T}}\boldsymbol{\pi}_E\right\}\end{bmatrix}$$

where $\boldsymbol{\pi}_E=-\boldsymbol{M}^{-1}\boldsymbol{r}$. The matrix $\boldsymbol{M}$ and the vector $\boldsymbol{r}$ are

$$\boldsymbol{M}=K\int_{t_0}^{t_f}e^{\boldsymbol{A}(t_f-t)}\boldsymbol{b}\boldsymbol{b}^{\mathrm{T}}\left(e^{\boldsymbol{A}(t_f-t)}\right)^{\mathrm{T}}\mathrm{d}t$$

$$\boldsymbol{r}=K\int_{t_0}^{t_f}e^{\boldsymbol{A}(t_f-t)}\boldsymbol{b}u\,\mathrm{d}t-\boldsymbol{K}_{\boldsymbol{g}_E}\boldsymbol{x}(t_f)+e^{\boldsymbol{A}(t_f-t_0)}\boldsymbol{K}_{x_0}\left(\boldsymbol{x}(t_0)-\begin{bmatrix}1\\1\end{bmatrix}\right)+\int_{t_0}^{t_f}e^{\boldsymbol{A}(t-s)}\boldsymbol{K}_f\left(\frac{\partial\boldsymbol{x}}{\partial t}(s)-\boldsymbol{A}\boldsymbol{x}(s)-\boldsymbol{b}u(s)\right)\mathrm{d}s$$

The one-dimensional matrix $K$ was set as $K=0.1$. The $2\times2$ dimensional matrixes $\boldsymbol{K}_{x_0}$, $\boldsymbol{K}_f$ and $\boldsymbol{K}_{\boldsymbol{g}_E}$ were both $\begin{bmatrix}0.1&0\\0&0.1\end{bmatrix}$. The definite conditions of the MEPDEs were directly set as $\left.\begin{bmatrix}x_1(t,\tau)\\x_2(t,\tau)\end{bmatrix}\right|_{\tau=0}=\begin{bmatrix}0.5\\0.5\end{bmatrix}$ and $u(t,\tau)|_{\tau=0}=0$. Using the semi-discrete method, the time horizon $[t_0,\,t_f]$ was discretized uniformly with 41 points. Thus, a dynamic system with 123 states was obtained and the OCP was transformed to a finite-dimensional IVP. Regarding the time derivative $\frac{\partial\boldsymbol{x}}{\partial t}$, it is computed with the finite difference method at the discretization points. The ODE integrator "ode45" in Matlab, with default relative error tolerance $1\times10^{-3}$ and default absolute error tolerance $1\times10^{-6}$, was employed to solve the IVP. For comparison, the analytic solution by solving the BVP is also presented.

$$\begin{aligned}\hat{x}_1 &= 0.5t^3 - 1.75t^2 + t + 1\\ \hat{x}_2 &= 1.5t^2 - 3.5t + 1\\ \hat{\lambda}_1 &= 3\\ \hat{\lambda}_2 &= -3t + 3.5\\ \hat{u} &= 3t - 3.5\end{aligned}$$

Figs. 2, 3 and 4 show the evolving process of $x_1(t)$, $x_2(t)$ and $u(t)$ solutions to the optimal, respectively. At $\tau$ = 300s, the numerical solutions are indistinguishable from the optimal, and this shows the effectiveness of the MEPDEs. In particular, it is found that the initial and boundary conditions are asymptotically met, as we expected. Fig. 5 plots the profile of performance index value against the variation time. Since starting from a zero solution, its value increases from zero, and then monotonously approaches the minimum, and it almost reaches the analytic value of 3.25 after $\tau$ =100s. In Fig. 6, the evolution profiles of Lagrange multipliers are presented. At $\tau$ =300s, we computed that $\boldsymbol{\pi}_E = \begin{bmatrix} 2.9963 \\ -2.4963 \end{bmatrix}$. From the analytic relations to the costates, we have

$$\boldsymbol{\lambda}(t) = \boldsymbol{\Phi}_o^{\mathrm{T}}(t_f, t)\boldsymbol{\pi}_E = \left(e^{\boldsymbol{A}(t_f - t)}\right)^{\mathrm{T}} \boldsymbol{\pi}_E = \begin{bmatrix} 1 & 0 \\ t_f - t & 1 \end{bmatrix}\begin{bmatrix} 2.9963 \\ -2.4963 \end{bmatrix} = \begin{bmatrix} 2.9963 \\ -2.9963t + 3.4963 \end{bmatrix}$$

This is very close to the analytic solution of $\begin{bmatrix} \hat{\lambda}_1 \\ \hat{\lambda}_2 \end{bmatrix}$.

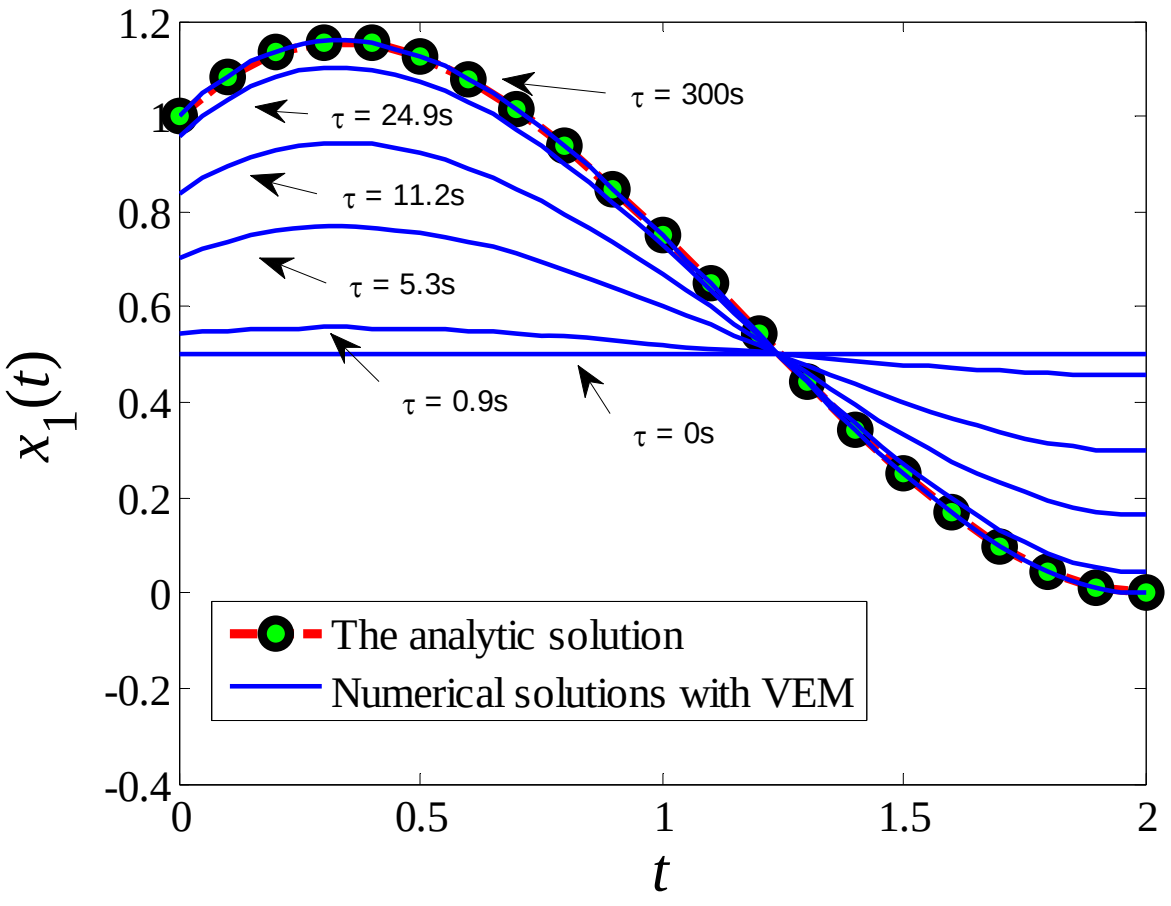

Fig. 2 The evolution of numerical solutions of $x_1$ to the optimal solution.

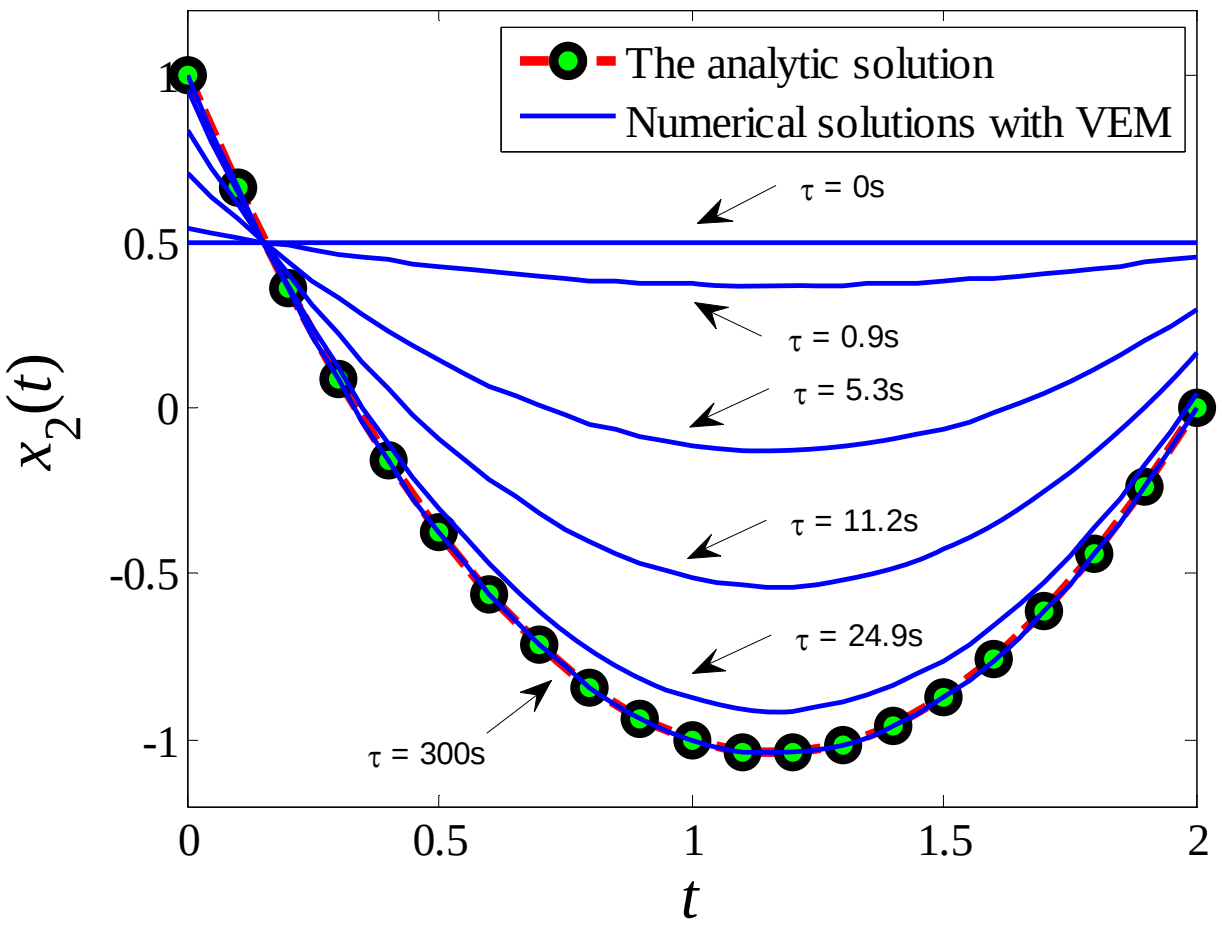

Fig. 3 The evolution of numerical solutions of $x_2$ to the optimal solution.

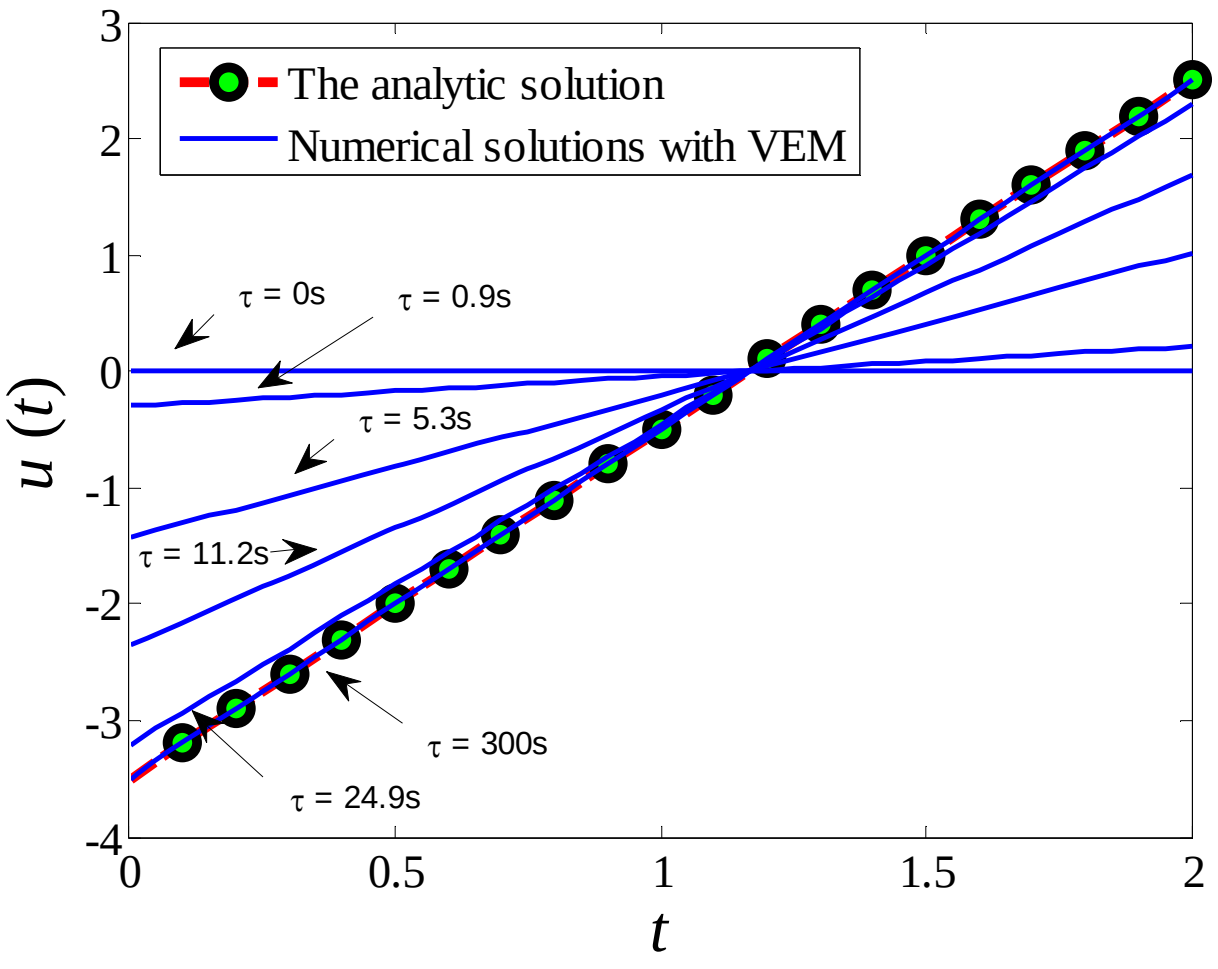

Fig. 4 The evolution of numerical solutions of $u$ to the optimal solution.

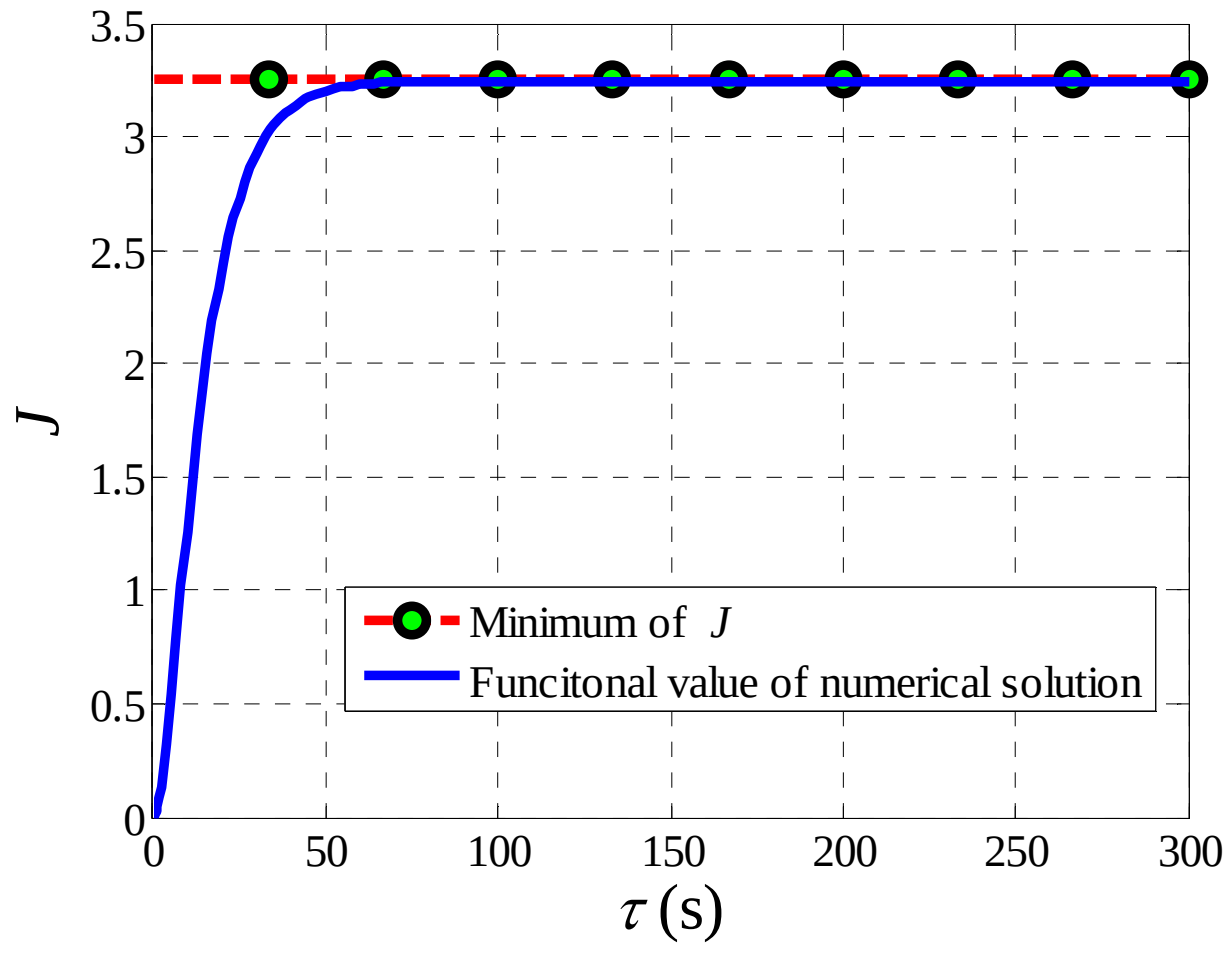

Fig. 5 The approach to the minimum of performance index.

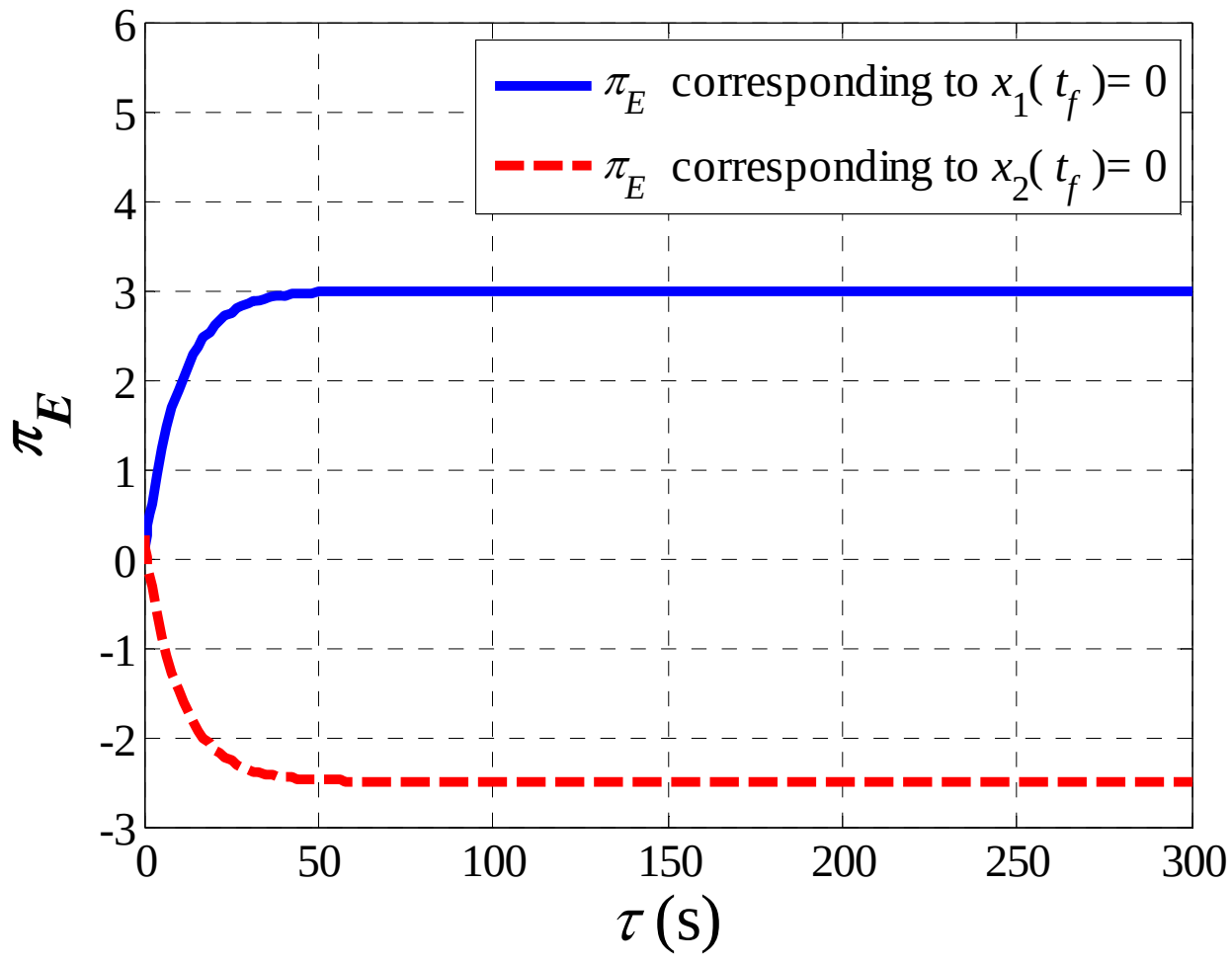

Fig. 6 The evolution profiles of Lagrange multipliers.

Now we consider a nonlinear example adapted from the Brachistochrone problem [24], which describes the motion curve of the fastest descending.

**Example 2**: Consider the following dynamic system

$$\dot{\boldsymbol{x}} = \boldsymbol{f}(\boldsymbol{x}, u)$$

where $\boldsymbol{x} = \begin{bmatrix} x \\ y \\ V \end{bmatrix}$, $\boldsymbol{f} = \begin{bmatrix} V\sin(u) \\ -V\cos(u) \\ g\cos(u) \end{bmatrix}$, $g = 10$ is the gravity constant. Find the solution that minimizes the performance index

$$J = t_f$$

with the boundary conditions

$$\left.\begin{bmatrix} x \\ y \\ V \end{bmatrix}\right|_{t_0=0} = \begin{bmatrix} 0 \\ 0 \\ 0 \end{bmatrix}, \quad \begin{matrix} x(t_f) = 2 \\ y(t_f) \le -2 \end{matrix}$$

In the specific form of the MEPDEs (9), (90) and the MEDE (91), the parameters $K$ and $k_{t_f}$ were set to be 0.1 and 0.05, respectively. The $3\times3$ dimensional matrixes $\boldsymbol{K}_{x_0}$ and $\boldsymbol{K}_f$ were both $\begin{bmatrix} 0.1 & 0 & 0 \\ 0 & 0.1 & 0 \\ 0 & 0 & 0.1 \end{bmatrix}$. The one-dimensional matrix $K_{\boldsymbol{g}_E}$ and the scalar parameter $k_{g_I}$ were both 0.1. The infeasible definite conditions, which satisfies the controllability condition, were set to be $\boldsymbol{x}(t,\tau)\big|_{\tau=0} = \begin{bmatrix} 1 \\ 1 \\ 1 \end{bmatrix}$ and $u(t,\tau)\big|_{\tau=0} = t$ with $t_f(\tau)\big|_{\tau=0} = 1s$. We also discretized the time horizon $[t_0, t_f]$ uniformly, with 101 points. Thus, a large IVP with 405 states (including the terminal time) was obtained. We still employed “ode45” in Matlab for the numerical integration. In the integrator setting, the default relative error tolerance and the absolute error tolerance were $1\times10^{-3}$ and $1\times10^{-6}$, respectively. For comparison, we computed the optimal solution with GPOPS-II [25], a Radau PS method based OCP solver.

Fig. 7 gives the states curve in the $xy$ coordinate plane, showing that the numerical results starting from the point $\begin{bmatrix}1\\1\end{bmatrix}$ approach the optimal solution over time. The control solutions are plotted in Fig. 8, and the asymptotical approach (with small oscillation) of the numerical results is demonstrated. In Fig. 9, the terminal time profile against the variation time $\tau$ is plotted. The result of $t_f$ declines rapidly at first and then gradually approaches the minimum decline time, and it only changes slightly after $\tau$ = 40s. At $\tau$ = 300s, we compute that $t_f$ = 0.8165s from the VEM, same to the result from GPOPS-II. Fig. 10 presents the evolution profiles of the Lagrange multiplier $\boldsymbol{\pi}_E$ and the KKT multiplier $\boldsymbol{\pi}_I$ . In particular, it is shown that $\boldsymbol{\pi}_I$ increases continuously from zero from $\tau$ = 4.2s, instead of changing suddenly. This phenomenon corresponds to the gradual convergence of $y(t_f)$ to the allowed value of -2, as shown in Fig. 11. At $\tau$ = 300s, we have $\boldsymbol{\pi}_E$ = -0.1477 and $\boldsymbol{\pi}_I$ =0.0564, same to the results in Ref. [15].

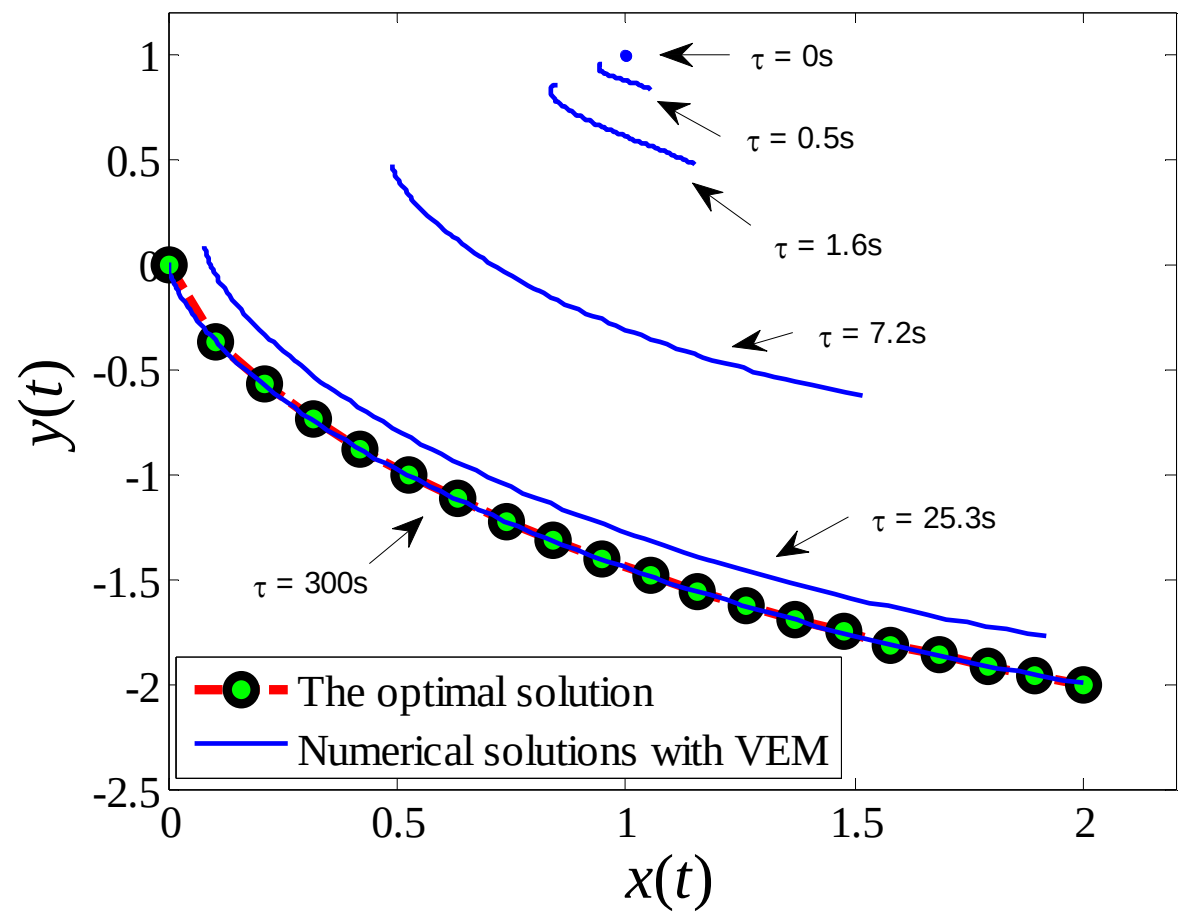

Fig. 7 The evolution of numerical solutions in the $xy$ coordinate plane to the optimal solution.

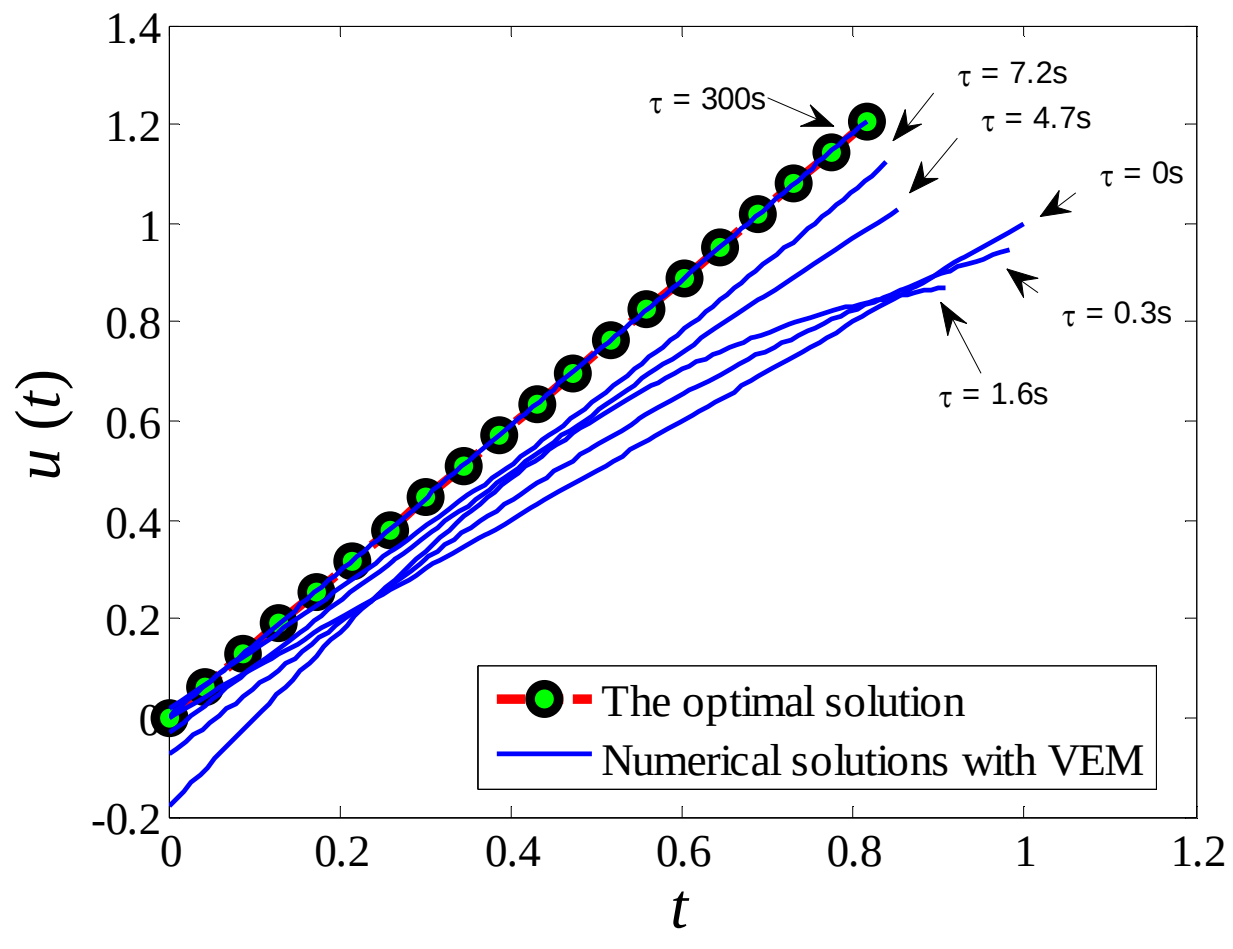

Fig. 8. The evolution of numerical solutions of $u$ to the optimal solution.

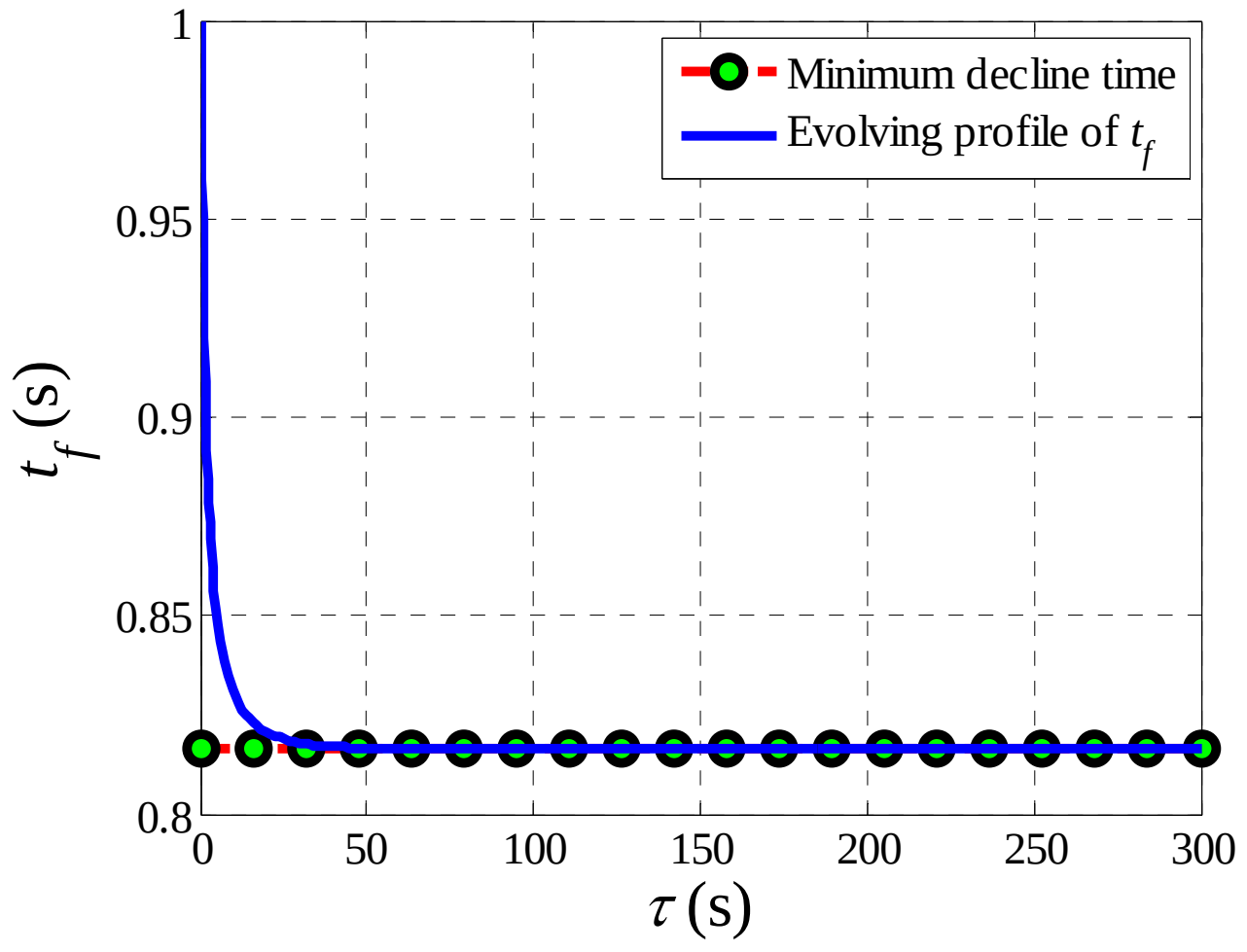

Fig. 9 The evolution profile of $t_f$ to the minimum decline time.

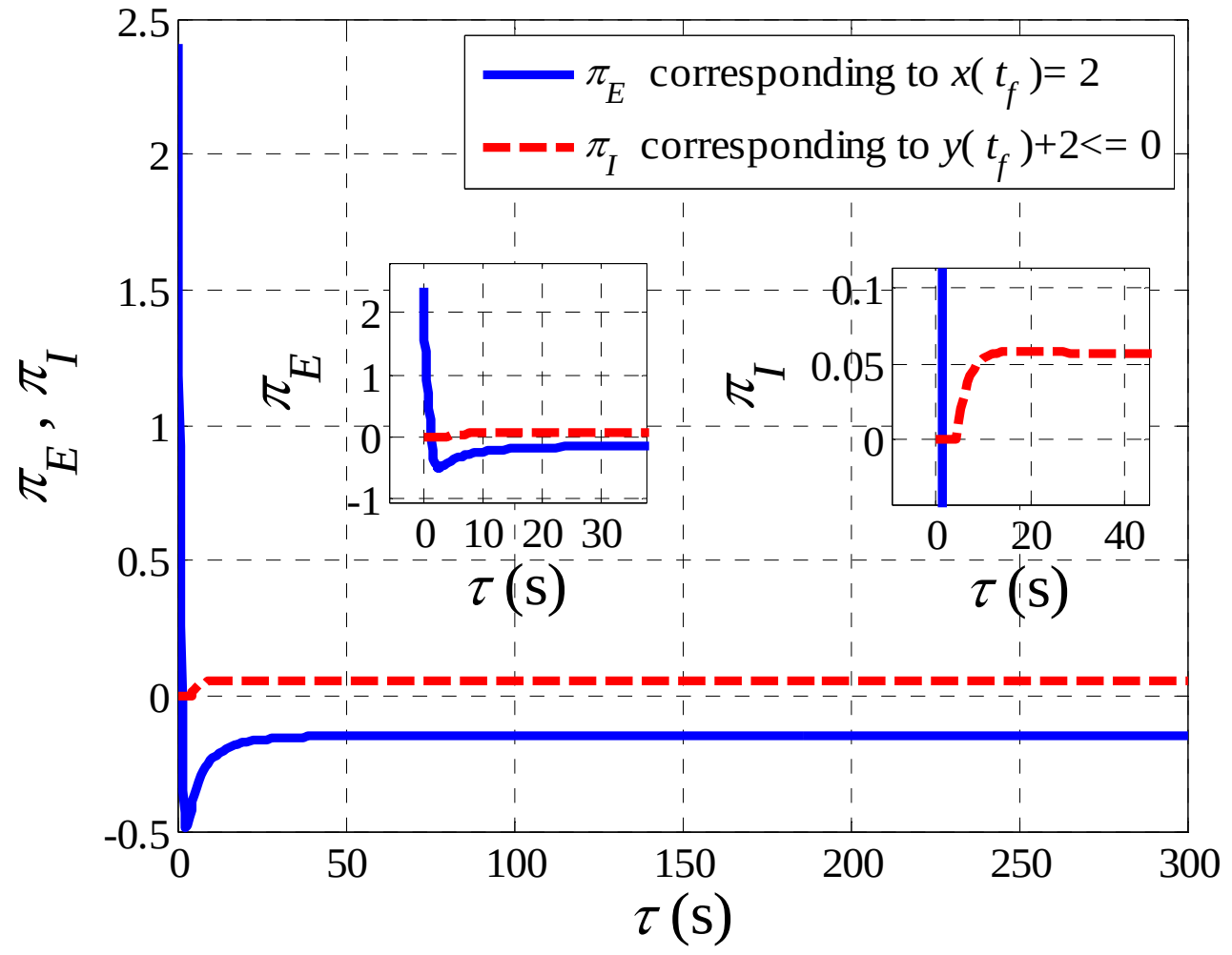

Fig. 10 The evolution profiles of Lagrange multiplier $\boldsymbol{\pi}_E$ and KKT multiplier $\boldsymbol{\pi}_I$ .

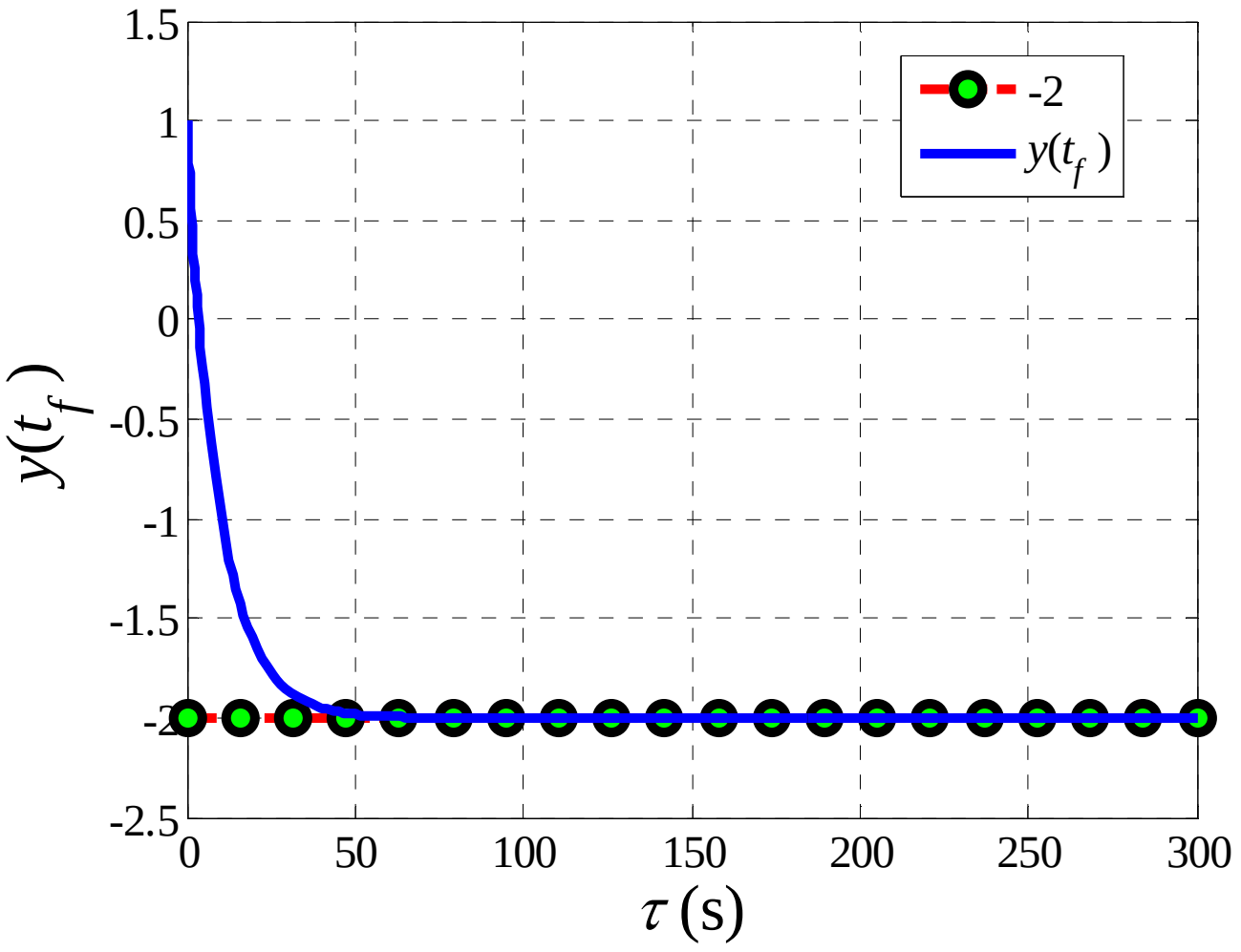

Fig. 11 The evolution profile of terminal IEC on $y(t_f)$ .

In order to inspect the finite time entry towards the feasible domain of IECs that are inactive for the optimal solution, we consider another version of this example, in which the terminal boundary conditions are specified as

$$x(t_f) = 2$$
$$y(t_f) \le -1$$

All the settings in the computation are still the same. Fig. 12 gives the evolution profile of the terminal IEC and the corresponding KKT multiplier, showing that $y(t_f)$ has entered the feasible domain since $\tau$ = 12.6s and $\pi_I$ always maintains zero during the evolution.

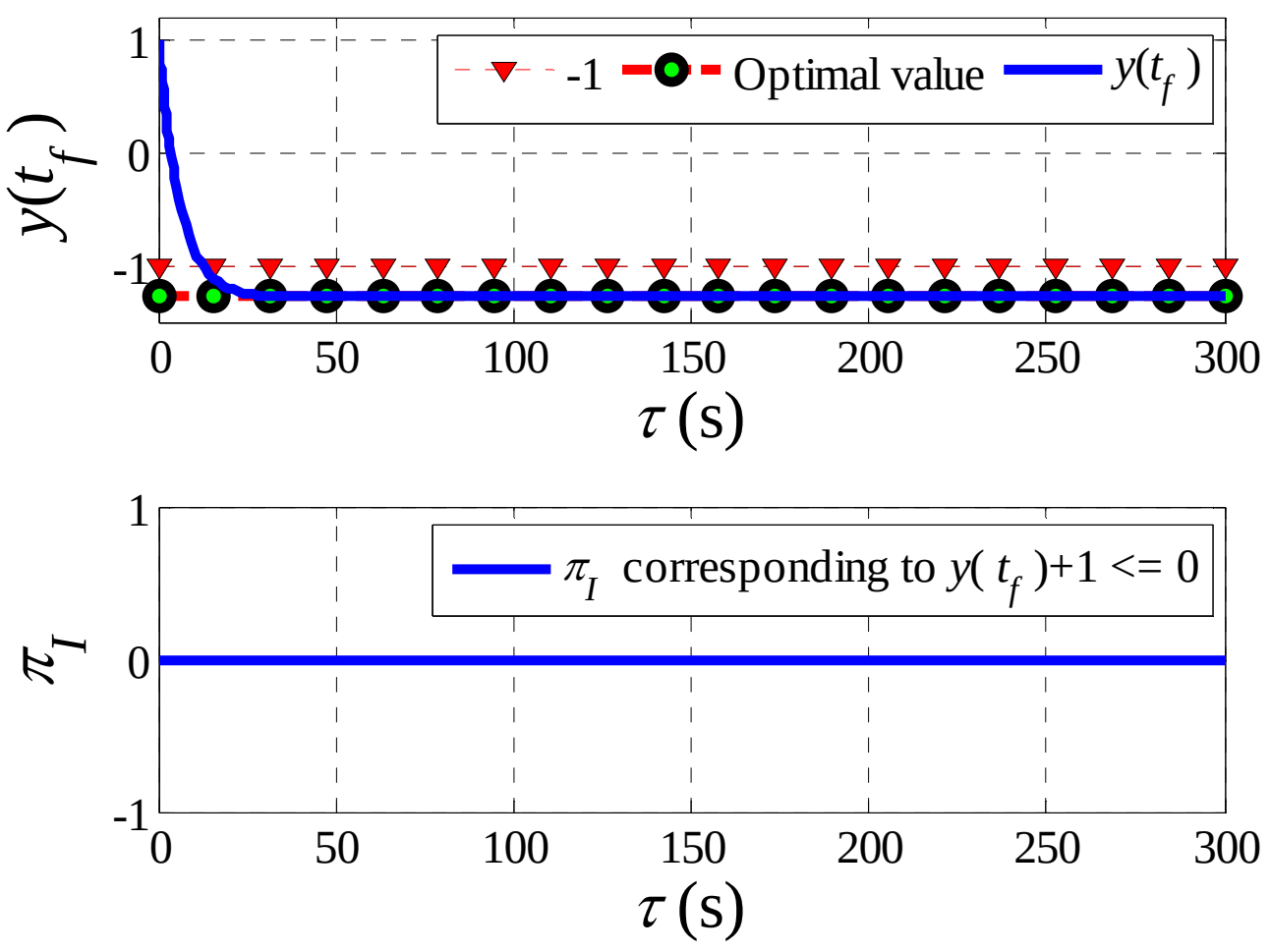

Fig. 12 The evolution profiles of terminal IEC on $y(t_f)$ and corresponding KKT multiplier.

## V. Conclusion

The Variation Evolving Method (VEM) is developed to be more flexible in solving the Optimal Control Problems (OCPs) with terminal (equality and inequality) constraint. Since the proposed Modified Evolution Partial Differential Equation (MEPDE) is valid even in the infeasible solution domain, in seeking the optimal solution, the requirement for a feasible definite condition is not necessary any more, and the transformed Initial-value Problems (IVPs) may be initialized with arbitrary initial values of variables. This brings great convenience because finding a feasible initial solution is usually not an easy task for the OCPs with terminal constraint. Numerical examples show that the MEPDE can effectively handle various infeasibilities, including the violated terminal inequality constraints. However, for the evolution from an infeasible solution, one must pay special attention to preserving the controllability of the dynamic system. If this is not satisfied, it may fail to generate the right solution.